# Graph-based sequential beamforming

Yongsung Park,[a] Florian Meyer,[b] and Peter Gerstoft

*Scripps Institution of Oceanography, University of California San Diego, La Jolla, California 92093-0238, USA*

**ABSTRACT:**
This paper presents a Bayesian estimation method for sequential direction finding. The proposed method estimates the number of directions of arrivals (DOAs) and their DOAs performing operations on the factor graph. The graph represents a statistical model for sequential beamforming. At each time step, belief propagation predicts the number of DOAs and their DOAs using posterior probability density functions (pdfs) from the previous time and a different Bernoulli-von Mises state transition model. Variational Bayesian inference then updates the number of DOAs and their DOAs. The method promotes sparse solutions through a Bernoulli-Gaussian amplitude model, is gridless, and provides marginal posterior pdfs from which DOA estimates and their uncertainties can be extracted. Compared to nonsequential approaches, the method can reduce DOA estimation errors in scenarios involving multiple time steps and time-varying DOAs. Simulation results demonstrate performance improvements compared to state-of-the-art methods. The proposed method is evaluated using ocean acoustic experimental data.
© 2023 Acoustical Society of America. https://doi.org/10.1121/10.0016876

## I. INTRODUCTION

Direction of arrival (DOA) estimation or beamforming[1] is the task of localizing several sources from noisy measurements provided by an array of sensors. DOA estimation is a key problem in electromagnetic, acoustic, and seismic applications. In this paper, we consider a scenario where multiple snapshots (or time steps) of data are available and the number of sources and their DOAs are time-varying. We propose an estimation method based on belief propagation (BP) and variational Bayesian inference [also known as the mean field (MF) approximation], which processes the data of dynamically varying DOAs sequentially. The proposed method is derived by applying message passing operations to the factor graph describing the inference problem's statistical model. BP is used to predict the number of sources and their DOAs based on a new state transition model and by using the previous time step's posterior probability density functions (pdfs). In addition, variational message passing is used to update DOAs and their number in an iterative update step. The resulting combined message passing method provides approximate marginal posterior pdfs that can be used to estimate the number of sources and their DOAs. A further advantage of using the framework of factor graphs is that it provides a graphical description of all of the algorithmic operations and shows the evolution of information over time.

### A. State-of-the-art

In most traditional DOA estimation methods (see Refs. 1 and 2 and references therein), a grid of potential DOAs and corresponding "steering vectors" is created to avoid nonlinear estimation. Of particular interest in recent years have been sparse DOA estimation methods that have high-resolution capabilities. For example, compressed sensing (CS)[3–6] approaches are based on a convex optimization procedure that explicitly promotes sparse solutions. Contrary to eigenvalue-based traditional DOA methods,[1,2] CS for DOA estimation[7,8] also performs well in scenarios with coherent sources.[3]

Another Bayesian approach[9,10] for DOA estimation is sparse Bayesian learning (SBL).[11–17] In SBL, a hierarchical prior model controls the prior variance of steering vector amplitudes and implicitly promotes sparse solutions. SBL-based methods can provide high DOA estimation accuracy but are known to overestimate the model order, i.e., they typically provide spurious low-power sources. An overview and discussion of grid-based CS and SBL type methods and insight into their differences are provided in Ref. 18. The main limitations of grid-based DOA estimation methods are (i) basis mismatch,[19–22] which is observed when look directions of steering vectors do not appropriately represent DOAs, and (ii) basis coherence,[7,13] which is caused by a dense grid of steering vectors and results in biased estimates.

In many scenarios, multiple snapshots of measurements are available, and the estimation of time-varying DOAs is performed across multiple time steps.[8,23–27] Here, sequential processing,[28–35] where information from previous times is used to compute DOA estimates at the current time, can improve the overall estimation performance. Sequential processing of sparse signals has been considered in Refs. 36–40, and sequential sparse DOA estimation has been developed based on CS[23] and SBL.[24,25,41] The main idea of these methods is to determine the parameters of sparsity-promoting prior pdfs using statistical information from previous times.

[a] Electronic mail: yongsungpark@ucsd.edu
[b] Also at: Department of Electrical and Computer Engineering, University of California San Diego, La Jolla, CA 92093-0238, USA.

Among emerging gridless sparse methods that are suitable for direction finding,[19,20,26,42–45] the variational Bayesian line spectral estimation (VALSE)[46–49] explicitly promotes sparsity by means of a Bernoulli-Gaussian amplitude model. It has favorable properties such that it (i) is guaranteed to converge, (ii) incorporates prior information, (iii) provides posterior pdfs of DOAs, and (iv) is gridless. Contrary to our approach, all of the existing methods are nonsequential processing. Variations of the VALSE are found for multi-snapshot processing[47] and potential two-dimensional (2-D) beamforming.[49] Multifrequency processing has not been explored.

### B. Contribution and notation

We propose a message passing method for sparse and gridless estimation of time-varying DOAs. Our approach uses the Bernoulli-Gaussian amplitude model of the VALSE[46–49] to promote sparsity and develops it with sequential estimation. Sequential Bayesian estimation is enabled by combining the amplitude model with a Bernoulli-von Mises state transition function. The resulting statistical model is described by a factor graph that provides the basis for developing the proposed message passing method for sequential variational Bayesian line spectral estimation (SVALSE).

The proposed method combines variational message passing (also referred to as the MF approximation[50–55]) and BP.[34,35,56–62] We use the theoretical framework in Ref. 63 to run MF on certain parts of the factor graph and BP on the others. MF approximates the joint posterior pdf with a factorization of marginal posterior pdfs. The marginal posterior pdfs are obtained by iteratively minimizing the Kullback-Leibler (KL) divergence of the joint posterior pdf with respect to the marginal posterior pdfs.[50–55] Iterative minimization is guaranteed to converge and can be interpreted as passing messages along the edges of the factor graph that represents the statistical model.[50–55] BP aims at computing marginal posterior pdfs from the joint posterior pdf in an efficient and scalable way. It operates by iteratively passing messages along the edges of a factor graph.[56–61] In case the factor graph is a tree, i.e., has no cycles, BP can provide the true joint posterior pdfs.[56–61] While BP yields an accurate approximation of marginal pdfs if the factor graph has no short cycles, MF always admits a convergent implementation and has closed-form message passing update rules for conjugate-exponential models.[63] The theoretical background for combining BP and MF on a single factor graph is found in Ref. 63.

We consider a Bernoulli-von Mises state transition model[24] and sparsity-promoting Bernoulli-Gaussian amplitude model[46] for sequential estimation of time-varying DOAs. To perform estimation efficiently, we leverage the benefits of BP and MF. In particular, we use BP for DOA prediction, where the proposed Bernoulli-von Mises state transition model leads to closed-form messages. BP cannot provide closed-form messages for DOA update. Thus, we use MF, which exploits the conjugate-exponential form of the amplitude model for efficient message computation. The resulting SVALSE method is an instance of the combined MF and BP, which is introduced in Ref. 63. Compared to nonsequential methods, our approach can achieve improved performance through a probabilistic information transfer between time steps. Contrary to existing sequential methods, the used factor graph formulation makes it possible to represent algorithmic operations and information propagation over time visually.

The key contributions of this paper are as follows:

(1) We establish a sparsity-promoting statistical model for sequential Bayesian estimation of time-varying DOAs and present the corresponding factor graph.
(2) We derive a Bayesian estimation method that computes approximate marginal pdfs efficiently by performing MF and BP message passing on the factor graph.
(3) We compare the performance of the proposed MF-BP message passing method with the state-of-the-art and demonstrate performance improvements.
(4) We validate our method using real acoustic measurements collected during the shallow water evaluation cell experiment 1996 (SwellEx-96), an underwater source localization experiment.

This paper advances the conference paper[24] by providing a detailed derivation of the proposed message passing method; discussing implementation aspects, including a new initialization; and presenting additional simulation and data processing results.

In what follows, random variables are displayed in sans serif, upright fonts (e.g., x) and their realizations are displayed in serif, italic (e.g., $x$). Vectors appear in bold lowercase (e.g., random, **x**, and realization, $\boldsymbol{x}$) and matrices appear in uppercase bold. Further, $f(\boldsymbol{x})$ denotes the pdf of continuous random vector **x** [short for $f_\mathbf{x}(\boldsymbol{x})$], $p(\boldsymbol{s})$ denotes the probability mass function (pmf) of discrete random vector **s** [short for $p_\mathbf{s}(\boldsymbol{s})$]. The expectation operator with respect to pdf $f(\boldsymbol{x})$ is given by $\mathsf{E}_{f(\boldsymbol{x})}[\cdot] = \int \cdot f(\boldsymbol{x}) \mathrm{d}\boldsymbol{x}$, similarly for pmf $p(\boldsymbol{s})$, $\mathsf{E}_{p(\boldsymbol{s})}[\cdot] = \sum_s \cdot p(\boldsymbol{s})$. Further, $\delta(x)$ is the Dirac delta function of continuous variable, $x$, and $\delta(s)$ is the Kronecker delta function of discrete variable, $s$. The complex multivariate Gaussian, $f_{\mathrm{CN}}(\cdot; \boldsymbol{\mu}, \boldsymbol{\Sigma})$, has mean $\boldsymbol{\mu}$ and covariance $\boldsymbol{\Sigma}$. The identity matrix of dimension $M$ is $\boldsymbol{I}_M$.

## II. SYSTEM MODEL AND PROBLEM FORMULATION

### A. Measurement model

We observe narrowband signals from $K_t$ sources with frequency, $\omega$, on an array of $M$ sensors. Let $c$ and $d \leq c\pi/\omega$ be propagation speed and sensor spacing, respectively.

At time $t \in \{0, 1, \ldots\}$, the received time-sampled signal, $\mathbf{y}_t = [\mathsf{y}_{0,t} \cdots \mathsf{y}_{M-1,t}]^T \in \mathbb{C}^M$, consists of elements

$$\mathsf{y}_{m,t} = \sum_{k=1}^{\mathsf{K}_t} \alpha_{k,t} e^{-jm(\omega d/c)\sin \beta_{k,t}^\star} + \mathsf{u}_{m,t}, \tag{1}$$

where $\beta_{k,t}^\star \in [-90°, 90°)$ and $\alpha_{k,t} \in \mathbb{C}$ are the true angle of arrival and the complex amplitude, respectively, of the signal component originated by source $k$, and $\mathsf{u}_{m,t} \in \mathbb{C}$ is additive noise. Note that at each time $t$, the complex amplitudes, $\alpha_{k,t}$, and angles, $\beta_{k,t}^\star$, $k \in \{1, \ldots, \mathsf{K}_t\}$, as well as the number, $\mathsf{K}_t$, of components are unknown. Estimating the number of components, $\mathsf{K}_t$, is typically referred to as model-order selection.[1,9]

Due to the unknown number of components, $\mathsf{K}_t$, we follow the approach in Ref. 46 and consider at most $L$ potential sources ($L > \mathsf{K}_t$). Each potential source corresponds to a random angle of arrival and weight. We formulate the underlying problem as line spectral estimation[22] by introducing pseudo angles (PAs), $\theta_{l,t} = -(\omega d/c)\sin \beta_{l,t} \in \Pi \triangleq [-\pi, \pi)$, and model the measurement vector, $\mathbf{y}_t$, as a realization of a random process,

$$\mathbf{y}_t = \sum_{l=1}^{L} \mathsf{w}_{l,t} \mathbf{a}(\theta_{l,t}) + \mathbf{u}_t, \tag{2}$$

where $\mathbf{a}(\theta_{l,t}) \triangleq [1 e^{j\theta_{l,t}} \cdots e^{j(M-1)\theta_{l,t}}]^T$ is the steering vector. We assume that the components of the noise, $\mathbf{u}_t \in \mathbb{C}^M$, are independent and identically distributed complex Gaussian with mean zero and variance $\nu$. For future reference, we introduce the vector $\mathbf{w}_t \triangleq [\mathsf{w}_{1,t} \cdots \mathsf{w}_{L,t}]^T \in \mathbb{C}^L$, the ordered sequences $\mathbf{w} \triangleq (\mathbf{w}_1, \ldots, \mathbf{w}_t)$, and $\mathbf{y} \triangleq (\mathbf{y}_1, \ldots, \mathbf{y}_t)$.

### B. Von Mises pdf

The von Mises pdf (VM; Ref. 64, p. 36) of angle $\theta \in [-\pi, \pi)$ is defined as

$$f_{\text{VM}}(\theta; \mu, \kappa) = \frac{1}{2\pi I_0(\kappa)} e^{\kappa \cos(\theta - \mu)}, \tag{3}$$

where $\mu$ and $\kappa$ are the mean direction and concentration, respectively, and $I_p(\cdot)$ is the modified Bessel function of the first kind and order $p$. Alternatively, the VM [Eq. (3)] can also be parametrized by $\eta = \kappa e^{j\mu}$ and then reads

$$f_{\text{VM}}(\theta; \eta) = \frac{1}{2\pi I_0(|\eta|)} \exp\bigl(\text{Re}\{\eta^H e^{j\theta}\}\bigr). \tag{4}$$

If $\kappa > 0$, the VM is symmetric around $\mu$ and has a similar shape as the Gaussian pdf. If $\kappa = 0$, the VM is uniform, i.e., $f(\theta) = 1/2\pi$. For large $\kappa$, the VM can be approximated accurately by a Gaussian with mean $\mu$ and variance $\sigma^2 = 1/\kappa$.

An important property of VMs is that they are closed under multiplication, i.e.,

$$f_{\text{VM}}(\theta; \eta_1) f_{\text{VM}}(\theta; \eta_2) \propto f_{\text{VM}}(\theta; \eta), \tag{5}$$

where $\eta = \eta_1 + \eta_2$. Thus, the resulting VM has mean direction $\arg(\eta_1 + \eta_2)$ and concentration $|\eta_1 + \eta_2|$.

### C. Prior pdfs

At each time $t$, only $\mathsf{K}_t$ of the $L$ components have non-zero weights. We use a sparsity-promoting prior for the complex weights, $\mathsf{w}_{l,t}$, which are governed by independent Bernoulli variables, $\mathsf{s}_{l,t} \in \mathcal{B} \triangleq \{0, 1\}$, i.e.,

$$f(\mathsf{w}_{l,t}|\mathsf{s}_{l,t}) = (1 - \mathsf{s}_{l,t})\delta(\mathsf{w}_{l,t}) + \mathsf{s}_{l,t} f_{\text{CN}}(\mathsf{w}_{l,t}; 0, \tau), \tag{6}$$

$$p(\mathsf{s}_{l,t}) = \rho_{l,t}^{\mathsf{s}_{l,t}}(1 - \rho_{l,t})^{(1-\mathsf{s}_{l,t})}. \tag{7}$$

Thus, the binary variable, $\mathsf{s}_{l,t}$, "deactivates" the $l$th component, i.e., $\mathsf{s}_{l,t} = 0$ implies that $\mathsf{w}_{l,t}$ is not used. If $\mathsf{s}_{l,t} = 1$, the $l$th component is active and $\mathsf{w}_{l,t}$ is zero-mean Gaussian with variance $\tau$. We set the probability, $\rho_{1,t} = \cdots = \rho_{L,t} = \rho$, when we do not have prior information on $\rho_{l,t}$. We also introduce vectors $\boldsymbol{\theta}_t \triangleq [\theta_{1,t} \cdots \theta_{L,t}]^T \in \Pi^L$ and $\mathbf{s}_t \triangleq [\mathsf{s}_{1,t} \cdots \mathsf{s}_{L,t}]^T \in \mathcal{B}^L$, as well as ordered sequences $\mathbf{s} \triangleq (\mathbf{s}_1, \ldots, \mathbf{s}_t)$ and $\boldsymbol{\theta} \triangleq (\boldsymbol{\theta}_1, \ldots, \boldsymbol{\theta}_t)$.

It is assumed that given $\mathbf{s}$, then $\mathbf{w}$ is statistically independent of $\boldsymbol{\theta}$ and the entries of $\mathbf{w}$ are statistical independent across $l$ and $t$, i.e.,

$$f(\mathbf{w}|\mathbf{s}, \boldsymbol{\theta}) = f(\mathbf{w}|\mathbf{s}) = \prod_{t'=1}^{t} \prod_{l=1}^{L} f(\mathsf{w}_{l,t'}|\mathsf{s}_{l,t'}). \tag{8}$$

For future reference, we introduce the set formed by the indices of nonzero entries in $\mathbf{s}$, i.e.,

$$\mathcal{S} = \{1 \leq l \leq L | \mathsf{s}_l = 1\}. \tag{9}$$

The PAs, $\boldsymbol{\theta}_t$, and Bernoulli variables, $\mathbf{s}_t$, are assumed to evolve independently and according to a first-order Markov model.[28,29] Further, at time $t = 0$, they are assumed statistically independent across $l$ with prior pdf $f(\boldsymbol{\theta}_0, \mathbf{s}_0) = \prod_{l=1}^{L} f_{\text{VM}}(\theta_{l,0}) p(\mathsf{s}_{l,0})$. Thus, the prior pdf of $\boldsymbol{\theta}$ and $\mathbf{s}$, which includes the Bernoulli-von Mises transition model, reads

$$f(\boldsymbol{\theta}, \mathbf{s}) = \left[\prod_{l=1}^{L} f_{\text{VM}}(\theta_{l,1}) p(\mathsf{s}_{l,1})\right] \\ \times \prod_{l=1}^{L} \prod_{t'=2}^{t} f_{\text{VM}}(\theta_{l,t'}|\theta_{l,t'-1}) p(\mathsf{s}_{l,t'}|\mathsf{s}_{l,t'-1}). \tag{10}$$

Here, $f_{\text{VM}}(\theta_{l,t}|\theta_{l,t-1})$ is a von Mises transition pdf. At time $t = 0$, the PAs, $\boldsymbol{\theta}_t$, have an arbitrary von Mises pdf and are assumed statistically independent, i.e., $f(\boldsymbol{\theta}_1) = \prod_{l=1}^{L} f_{\text{VM}}(\theta_{l,1})$. The transition pmf, $p(\mathsf{s}_{l,t}|\mathsf{s}_{l,t-1})$, is

$$p(\mathsf{s}_{l,t}|\mathsf{s}_{l,t-1}) = \begin{cases} p^d & \text{if } \mathsf{s}_{l,t} = 0 \text{ and } \mathsf{s}_{l,t-1} = 1, \\ p^a & \text{if } \mathsf{s}_{l,t} = 1 \text{ and } \mathsf{s}_{l,t-1} = 0. \end{cases} \tag{11}$$

If component $l$ is active at time $t-1$, i.e., $\mathsf{s}_{l,t-1} = 1$, then the probability that it is inactive at time $t$, i.e., $\mathsf{s}_{l,t} = 0$, is given by the deactivation probability, $p^d$. Further, if component $l$ was inactive at time $t-1$, i.e., $\mathsf{s}_{l,t-1}=0$, then the probability that it is active at time $t$, i.e., $\mathsf{s}_{l,t}=1$, is given by the activation probability, $p^a$; see Sec. V.

### D. Likelihood function and joint posterior pdf

The measurement model [Eq. (2)] implies that given **w**, the measurements, **y**, are statistically independent of **s**. Further, the components of the noise $\mathbf{u}_t$ [Eq. (2)] are assumed independent and identically distributed (iid) complex zero-mean Gaussian with variance, $\nu$, and statistically independent across time, $t$. This results in the joint likelihood function as

$$f(\mathbf{y}|\boldsymbol{\theta},\mathbf{w},\mathbf{s}) = f(\mathbf{y}|\boldsymbol{\theta},\mathbf{w}) = \prod_{t'=1}^{t} f(\mathbf{y}_{t'}|\boldsymbol{\theta}_{t'},\mathbf{w}_{t'}), \quad (12)$$

$$f(\mathbf{y}_{t'}|\boldsymbol{\theta}_{t'},\mathbf{w}_{t'}) = f_{\mathrm{CN}}\left(\mathbf{y}_{t'}; \sum_{l=1}^{L} w_{l,t'}\mathbf{a}(\theta_{l,t'}), \nu \mathbf{I}_M\right). \quad (13)$$

#### 1. Joint pdfs

Using the chain rule, we obtain the joint pdf as

$$\begin{aligned}f(\boldsymbol{\theta},\mathbf{w},\mathbf{s},\mathbf{y}) &= f(\mathbf{y}|\boldsymbol{\theta},\mathbf{w},\mathbf{s})f(\boldsymbol{\theta},\mathbf{w},\mathbf{s})\\ &= f(\mathbf{y}|\boldsymbol{\theta},\mathbf{w},\mathbf{s})f(\mathbf{w}|\mathbf{s},\boldsymbol{\theta})f(\mathbf{s},\boldsymbol{\theta}).\end{aligned} \quad (14)$$

Inserting Eq. (12) for $f(\mathbf{y}|\boldsymbol{\theta},\mathbf{w},\mathbf{s})$, Eq. (8) for $f(\mathbf{w}|\mathbf{s},\boldsymbol{\theta})$, and Eq. (10) for $f(\mathbf{s},\boldsymbol{\theta})$, yields the following factorization of the joint pdf:

$$\begin{aligned}f(\boldsymbol{\theta},\mathbf{w},\mathbf{s},\mathbf{y}) = &\left[f(\mathbf{y}_1|\boldsymbol{\theta}_1,\mathbf{w}_1)\prod_{l=1}^{L}f(w_{l,1}|s_{l,1})f_{\mathrm{VM}}(\theta_{l,1})p(s_{l,1})\right]\\ &\times \prod_{t'=2}^{t} f(\mathbf{y}_{t'}|\boldsymbol{\theta}_{t'},\mathbf{w}_{t'})\prod_{l=1}^{L} f(w_{l,t'}|s_{l,t'})\\ &\times f(\theta_{l,t'}|\theta_{l,t'-1})p(s_{l,t'}|s_{l,t'-1}).\end{aligned} \quad (15)$$

Because the measurements, **y**, are observed and, thus, fixed, the joint posterior pdf is $f(\boldsymbol{\theta},\mathbf{w},\mathbf{s}|\mathbf{y}) \propto f(\boldsymbol{\theta},\mathbf{w},\mathbf{s},\mathbf{y})$.

In Fig. 1, we illustrate the factor graph[56] for the joint posterior pdf [Eq. (15)] in short notation, which is given by

$$f(\boldsymbol{\theta},\mathbf{w},\mathbf{s},\mathbf{y}) = \left[f_{\mathbf{y}|\boldsymbol{\theta}\mathbf{w}}^{(t=1)}\prod_{l=1}^{L}f_{\mathsf{w}_l|\mathsf{s}_l}^{(t=1)}f_{\theta_l}^{(t=1)}p_{\mathsf{s}_l}^{(t=1)}\right]$$
$$\times \prod_{t'=2}^{t} f_{\mathbf{y}|\boldsymbol{\theta}\mathbf{w}}^{(t=t')}\prod_{l=1}^{L}f_{\mathsf{w}_l|\mathsf{s}_l}^{(t=t')}f_{\theta_l}^{(t=t')}p_{l}^{(t=t')}, \quad (16)$$

where we specialize it to $t=2$ time steps as in Fig. 1,

$$f(\boldsymbol{\theta},\mathbf{w},\mathbf{s},\mathbf{y}) = f_{\mathbf{y}|\boldsymbol{\theta}\mathbf{w}}^{(t=1)}f_{\mathbf{y}|\boldsymbol{\theta}\mathbf{w}}^{(t=2)}$$
$$\times \prod_{l=1}^{L}f_{\mathsf{w}_l|\mathsf{s}_l}^{(t=1)}f_{\theta_l}^{(t=1)}p_{\mathsf{s}_l}^{(t=1)}f_{\mathsf{w}_l|\mathsf{s}_l}^{(t=2)}f_{l}^{(t=2)}p_{l}^{(t=2)}. \quad (17)$$

The factor for $t=1$ corresponds to the joint posterior pdf of nonsequential VALSE.[46,47] The factor graph in Fig. 1 serves

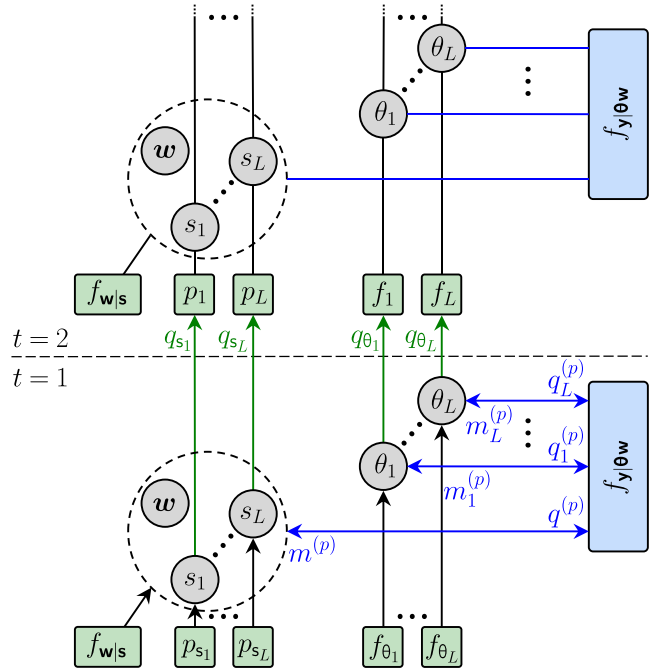

FIG. 1. (Color online) Factor graph for sequential Bayesian sparse DOAs estimation corresponding to the factorization [Eq. (15)]. Factor nodes representing prior pdfs and likelihood functions are shown as blue and green boxes, respectively. Variable nodes representing the random parameters are shown as circles. The time indices, $t$, are omitted in the notations of the nodes and messages, $m$, between factor nodes and nodes are indicated.

as the blueprint for the development of a message passing algorithm[56,63,65] in Secs. III and V.

### E. Problem formulation

We consider the problem of Bayesian detecting and estimating the DOAs, $\theta_{k,t}$, $k \in \{1,...,\mathsf{K}_t\}$, among the PAs, $\theta_{l,t}$, $l \in \{1,...,L\}$ ($L > \mathsf{K}_t$) from all of the observations, **y**, over time, $t$. This relies on the marginal posterior activation pmf, $p(s_{l,t}|\mathbf{y})$, and the marginal posterior pdfs, $f(\theta_{l,t}|\mathbf{y})$. A PA, $\theta_{l,t}$, is declared active if $p(s_{l,t}=1|\mathbf{y}) > P_{\mathrm{th}}$, where $P_{\mathrm{th}} \in [0,1)$ is a detection threshold (Ref. 66, Chap. 2). For PAs that are considered active, the PAs, $\theta_{l,t}$, are estimated by the minimum mean square error (MMSE) estimator (Ref. 66, Chap. 4) for angles,

$$\hat{\theta}_l = \arg\left(\mathsf{E}_{f(\theta_l|\mathbf{y})}[e^{j\theta_l}]\right). \quad (18)$$

From the PAs, $\hat{\theta}_{l,t}$, the estimated DOAs are $\hat{\beta}_{l,t} = \sin^{-1}(-(c/\omega d)\hat{\theta}_{l,t})$.

Calculation of the pdfs, $f(\theta_{l,t}|\mathbf{y})$, and pmfs, $p(s_{l,t}|\mathbf{y})$, needed for PA state detection and estimation by direct marginalization from the joint posterior, $f(\boldsymbol{\theta},\mathbf{w},\mathbf{s}|\mathbf{y})$, in Eq. (15) is infeasible. In our approach, for each $l \in \{1,...,L\}$ at time $t$, we sequentially predict and update approximations, $q(\theta_{l,t}|\mathbf{y})$ and $q(s_{l,t}|\mathbf{y})$, of the marginal pdfs, $f(\theta_{l,t}|\mathbf{y})$ and $p(s_{l,t}|\mathbf{y})$. Also, the model parameters, the noise variance $\nu$, and variance, $\tau$, of the Bernoulli-Gaussian prior model are updated. For the update, we consider a feasible approximate calculation by means of variational Bayesian estimation[50] based on VALSE.[46] Within this estimation approach, the resulting approximate marginal activation pmfs, $q(s_{l,t}|y)$, have the form $q(s_{l,t}|y) = \delta(s_{l,t} - \hat{s}_{l,t})$. Thus, the thresholding based on $P_{\text{th}}$ is avoided, i.e., PA $\theta_{l,t}$ is active at time $t$ if $\hat{s}_{l,t} = 1$.

## III. VARIATIONAL BAYESIAN APPROXIMATION

Here, we review the variational Bayesian approach (Ref. 50, Chap. 10; Ref. 51, Chap. 21; Ref. 52, Chap. 10) based on the system model in Sec. II. We focus on one time step, i.e., we use the joint posterior pdf [Eq. (15)] at $t = 1$. For simplicity, the time index, $t$, is omitted. The joint pdf now reads

$$f(\boldsymbol{\theta}, \boldsymbol{w}, \boldsymbol{s}, \boldsymbol{y}) \propto f(\boldsymbol{y}|\boldsymbol{\theta}, \boldsymbol{w}) \prod_{l=1}^{L} f(w_l|s_l) f_{\text{VM}}(\theta_l) p(s_l). \quad (19)$$

### A. Variational Bayes

The variational Bayesian approach aims to approximate $f(\boldsymbol{\Phi}|\boldsymbol{y})$ by a simpler pdf, $q(\boldsymbol{\Phi}|\boldsymbol{y})$, which is in a family of tractable pdfs. The parameters are collectively denoted by $\boldsymbol{\Phi} = \{\boldsymbol{\theta}, \boldsymbol{w}, \boldsymbol{s}\}$. Approximation is performed by minimizing the *KL divergence* (Ref. 50, p. 463; Ref. 51, p.732; Ref. 67, p. 205), i.e.,

$$\text{KL}[q(\boldsymbol{\Phi}|\boldsymbol{y})||f(\boldsymbol{\Phi}|\boldsymbol{y})] = -\mathsf{E}_{q(\boldsymbol{\Phi}|\boldsymbol{y})}\left[\ln \frac{f(\boldsymbol{\Phi}|\boldsymbol{y})}{q(\boldsymbol{\Phi}|\boldsymbol{y})}\right]. \quad (20)$$

Note that the KL divergence is non-negative (Ref. 67, Chap. 6.2.4). Using $f(\boldsymbol{\Phi}|\boldsymbol{y}) = f(\boldsymbol{\Phi}, \boldsymbol{y})/f(\boldsymbol{y})$ in Eq. (20), the log model evidence, $\ln f(\boldsymbol{y})$, is given by [Ref. 50, Eq. (10.2)] as

$$\ln f(\boldsymbol{y}) = \text{KL}[q(\boldsymbol{\Phi}|\boldsymbol{y})||f(\boldsymbol{\Phi}|\boldsymbol{y})] + \mathcal{L}[q(\boldsymbol{\Phi}|\boldsymbol{y})], \quad (21)$$

$$\mathcal{L}[q(\boldsymbol{\Phi}|\boldsymbol{y})] \triangleq \mathsf{E}_{q(\boldsymbol{\Phi}|\boldsymbol{y})}\left[\ln \frac{f(\boldsymbol{\Phi}, \boldsymbol{y})}{q(\boldsymbol{\Phi}|\boldsymbol{y})}\right]. \quad (22)$$

Since $\ln f(\boldsymbol{y}) \geq \mathcal{L}[q(\boldsymbol{\Phi}|\boldsymbol{y})]$, due to the nonnegativity of the KL divergence (Ref. 67, Chap. 6.2.4), $\mathcal{L}$ is the *evidence lower bound* (ELBO; Ref. 52, Chap. 10.1.2; Refs. 68 and 69). For observed and, thus, fixed measurements, $\boldsymbol{y} \in \mathbb{C}^M$, $\ln f(\boldsymbol{y})$ is a constant [Eq. (21)]. Minimizing the KL divergence [Eq. (20)] is, therefore, equivalent to maximizing the ELBO [Eq. (22)].

For the development of the VALSE method,[46] the ELBO, $\mathcal{L}[q(\boldsymbol{\Phi}|\boldsymbol{y})]$, in Eq. (22) is maximized by assuming that $q(\boldsymbol{\Phi}|\boldsymbol{y})$ is in the following family of tractable pdfs:

$$q(\boldsymbol{\Phi}|\boldsymbol{y}) = q(\boldsymbol{\theta}, \boldsymbol{w}, \boldsymbol{s}|\boldsymbol{y}) = q(\boldsymbol{w}, \boldsymbol{s}|\boldsymbol{y}) \prod_{l=1}^{L} q(\theta_l|\boldsymbol{y}), \quad (23)$$

where the factor $q(\boldsymbol{w}, \boldsymbol{s}|\boldsymbol{y}) = q(\boldsymbol{w}|\boldsymbol{y}, \boldsymbol{s})q(\boldsymbol{s}|\boldsymbol{y})$ is further constrained by setting $q(\boldsymbol{s}|\boldsymbol{y}) = \delta(\boldsymbol{s} - \hat{\boldsymbol{s}})$, i.e., the posterior pmf, $q(\boldsymbol{s}|\boldsymbol{y})$, of the binary vector, $\boldsymbol{s} \in \mathcal{B}^L$, has all of its mass at a single vector, $\hat{\boldsymbol{s}} \in \mathcal{B}^L$. Thus, the final expression for the considered family of tractable pdfs reads

$$q(\boldsymbol{\theta}, \boldsymbol{w}, \boldsymbol{s}|\boldsymbol{y}) = q(\boldsymbol{w}|\boldsymbol{y}, \boldsymbol{s})\delta(\boldsymbol{s} - \hat{\boldsymbol{s}}) \prod_{l=1}^{L} q(\theta_l|\boldsymbol{y}). \quad (24)$$

For future reference, $\hat{\mathcal{S}}$ is the set of indices of the nonzero entries in $\hat{\boldsymbol{s}}$ and due to $q(\boldsymbol{s}|\boldsymbol{y}) = \delta(\boldsymbol{s} - \hat{\boldsymbol{s}})$, the equalities hold,

$$q(\boldsymbol{w}|\boldsymbol{y}) = \sum_{\boldsymbol{s}} q(\boldsymbol{w}, \boldsymbol{s}|\boldsymbol{y}) = \sum_{\boldsymbol{s}} q(\boldsymbol{w}|\boldsymbol{y}, \boldsymbol{s})\delta(\boldsymbol{s} - \hat{\boldsymbol{s}}) = q(\boldsymbol{w}|\boldsymbol{y}, \hat{\boldsymbol{s}}).$$

$$(25)$$

## IV. VARIATIONAL LINE SPECTRAL ESTIMATION

A message passing algorithm for variational line spectral estimation is presented with closed-form expressions for the resulting message and approximate marginal pdfs.

Following the variational Bayesian estimation approach (Ref. 50, Chap. 10.1.1; Ref. 51, Chap. 21.3; Ref. 52, Chap. 10.2), an alternating optimization is performed because maximizing the ELBO, $\mathcal{L}[q(\boldsymbol{\theta}, \boldsymbol{w}, \boldsymbol{s}|\boldsymbol{y})]$, for all of the approximate marginal pdfs $q(\theta_l|\boldsymbol{y})$, $l = 1, \ldots, L$ and $q(\boldsymbol{w}, \boldsymbol{s}|\boldsymbol{y})$ simultaneously is infeasible. The ELBO, $\mathcal{L}[q(\boldsymbol{\theta}, \boldsymbol{w}, \boldsymbol{s}|\boldsymbol{y})]$, is maximized in turns over each of the approximate marginal pdfs $q(\boldsymbol{w}|\boldsymbol{y}, \boldsymbol{s})$ and $q(\theta_l|\boldsymbol{y})$, whereas the others are kept fixed. After initialization ($p=0$) at each iteration $p \in \{1, \ldots, P\}$, variational Bayesian estimation cycles through these approximate marginal pdfs and replaces them one by one with updated versions. Such an alternating optimization approach is guaranteed to converge to a local maximum of the ELBO, $\mathcal{L}[q(\boldsymbol{\theta}, \boldsymbol{w}, \boldsymbol{s}|\boldsymbol{y})]$ (Ref. 50, Chap. 10.1).

### A. MF message passing

Let $q^{(p-1)}(\theta_l|\boldsymbol{y})$ and $q^{(p-1)}(\boldsymbol{w}, \boldsymbol{s}|\boldsymbol{y})$ be the approximate marginal pdfs updated at the previous iteration $p-1$. At iteration $p$, first the updated approximate marginal pdf, $q^{(p)}(\theta_l|\boldsymbol{y})$, $l \in \{1, \ldots, L\}$, are obtained as [Ref. 50, Eq. (10.9); Ref. 51, Eq. (21.25)]

$$q^{(p)}(\theta_l|\boldsymbol{y}) \propto \exp\left(\mathsf{E}_{\sim q^{(p-1)}(\theta_l|\boldsymbol{y})}[\ln f(\boldsymbol{y}, \boldsymbol{\theta}, \boldsymbol{w}, \boldsymbol{s})]\right), \quad (26)$$

where $\mathsf{E}_{\sim q^{(p-1)}(\theta_l|\boldsymbol{y})}[\cdot]$ is the expectation with respect to $q^{(p)}(\boldsymbol{w}, \boldsymbol{s}|\boldsymbol{y})(\prod_{l'=1}^{l-1} q^{(p)}(\theta_{l'}|\boldsymbol{y})) \prod_{l'=l+1}^{L} q^{(p-1)}(\theta_{l'}|\boldsymbol{y})$. Equation (26) says that the optimal factor, $q(\theta_l|\boldsymbol{y})$, is obtained considering the joint pdf over all of the variables $f(\boldsymbol{y}, \boldsymbol{\theta}, \boldsymbol{w}, \boldsymbol{s})$ [Eq. (19)] and then taking the expectation with respect to all of the other factors except $q(\theta_l|\boldsymbol{y})$, i.e., $q(\boldsymbol{w}, \boldsymbol{s}|\boldsymbol{y})$ and $\{q(\theta_{l'}|\boldsymbol{y})\}$ for $l' \neq l$ [Eq. (23)]. Plugging $f(\boldsymbol{y}, \boldsymbol{\theta}, \boldsymbol{w}, \boldsymbol{s})$ in Eq. (19) into Eq. (26), we obtain

$$q^{(p)}(\theta_l|\boldsymbol{y}) \propto f_{\text{VM}}(\theta_l) m_l^{(p-1)}(\theta_l), \quad (27)$$

where we introduce the *messages*

$$m_l^{(p-1)}(\theta_l) \triangleq \exp\left(\mathsf{E}_{\sim q^{(p-1)}(\theta_l|\boldsymbol{y})}[\ln f(\boldsymbol{y}|\boldsymbol{\theta}, \boldsymbol{w})]\right), \quad (28)$$

where $f(\boldsymbol{y}|\boldsymbol{\theta}, \boldsymbol{w})$ is the likelihood [Eq. (12)]. At initialization ($p=0$), Eq. (27) is initialized as $q^{(0)}(\theta_l|\boldsymbol{y}) \propto f_{\text{VM}}(\theta_l)$, which is obtained using Eq. (66).

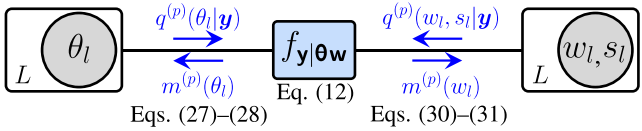

FIG. 2. (Color online) MF message passing procedure for variational inference. A portion of the factor graph in Fig. 1 appears. The box (plate) denotes a set of nodes, and the connections are duplicated $L$ times.

Similarly, as in Eq. (26), the approximate marginal pdfs $q^{(p)}(\mathbf{w}, \mathbf{s}|\mathbf{y})$ are updated, i.e.,

$$q^{(p)}(\mathbf{w}, \mathbf{s}|\mathbf{y}) \propto \exp\left(\mathsf{E}_{\sim q^{(p-1)}(\mathbf{w},\mathbf{s}|\mathbf{y})}[\ln f(\mathbf{y}, \boldsymbol{\theta}, \mathbf{w}, \mathbf{s})]\right), \quad (29)$$

where $\mathsf{E}_{\sim q^{(p-1)}(\mathbf{w},\mathbf{s}|\mathbf{y})}[\cdot]$ is the expectation with respect to $q^{(p-1)}(\boldsymbol{\theta}|\mathbf{y}) = \prod_{l=1}^{L} q(\theta_l|\mathbf{y})$. Plugging $f(\mathbf{y}, \boldsymbol{\theta}, \mathbf{w}, \mathbf{s})$ in Eq. (19) into Eq. (29), we obtain

$$q^{(p)}(\mathbf{w}, \mathbf{s}|\mathbf{y}) \propto f(\mathbf{w}|\mathbf{s}) p(\mathbf{s}) m^{(p-1)}(\mathbf{w}), \quad (30)$$

where $f(\mathbf{w}|\mathbf{s}) \triangleq \prod_{l=1}^{L} f(w_l|s_l)$ [Eq. (6)], $p(\mathbf{s}) \triangleq \prod_{l=1}^{L} p(s_l)$, and the messages,

$$m^{(p-1)}(\mathbf{w}) \triangleq \exp\left(\mathsf{E}_{\sim q^{(p-1)}(\mathbf{w},\mathbf{s}|\mathbf{y})}[\ln f(\mathbf{y}|\boldsymbol{\theta}, \mathbf{w})]\right). \quad (31)$$

At initialization $(p = 0)$, Eq. (30) is initialized with $q^{(0)}(\mathbf{w}, \mathbf{s}|\mathbf{y})$, which is obtained using Eq. (51) based on $q^{(0)}(\boldsymbol{\theta}|\mathbf{y})$ [Eq. (66)]. Equations (27) and (30) introduce a MF message passing algorithm.[65] MF messages passed along the edges of the factor graphs in Figs. 1 and 2 are shown in blue.

### B. Computing $q^{(p)}(\theta_l|\mathbf{y})$

Let $\hat{\mathbf{s}}$ be the activation vector from previous iteration $p-1$, and let $\hat{\mathcal{S}}$ be the set of corresponding nonzero entries in $\hat{\mathbf{s}}$; see Eq. (57). Computing the expectation $\mathsf{E}_{\sim q^{(p-1)}(\theta_l|\mathbf{y})}[\cdot]$ in Eq. (26) for $l \in \hat{\mathcal{S}}$, gives (Ref. 46, Sec. III A)

$$q^{(p)}(\theta_l|\mathbf{y}) \propto f_{\mathrm{VM}}(\theta_l) \exp\left\{\mathrm{Re}\left(\boldsymbol{\eta}_l^H \mathbf{a}(\theta_l)\right)\right\}, \quad (32)$$

where $\boldsymbol{\eta}_l \in \mathbb{C}^M$ is given by

$$\boldsymbol{\eta}_l = \frac{2}{\nu}\left(\mathbf{y} - \sum_{l' \in \hat{\mathcal{S}} \setminus \{l\}} \hat{w}_{l'} \hat{\mathbf{a}}_{l'}\right) \hat{w}_l^H - \frac{2}{\nu} \sum_{l' \in \hat{\mathcal{S}} \setminus \{l\}} \hat{C}_{l,l'} \hat{\mathbf{a}}_l. \quad (33)$$

Here, steering vectors, $\hat{\mathbf{a}}_{l'}$, are estimated from the approximate marginal pdf, $q^{(p)}(\theta_l|\mathbf{y})$, using the characteristic function (Ref. 64, Chap. 3.3) of the VM pdf (Ref. 64, Chap. 3.5.4),

$$\hat{\mathbf{a}}_{l'} \triangleq \begin{cases} \mathsf{E}_{q^{(p)}(\theta_{l'}|\mathbf{y})}[\mathbf{a}(\theta_{l'})], & l' \in \hat{\mathcal{S}} \cap \{1, ..., l-1\}, \\ \mathsf{E}_{q^{(p-1)}(\theta_{l'}|\mathbf{y})}[\mathbf{a}(\theta_{l'})], & l' \in \hat{\mathcal{S}} \cap \{l+1, ..., L\}. \end{cases} \quad (34)$$

Further, for $l \in \hat{\mathcal{S}}$, estimates of weights, $\hat{w}_l$, and corresponding variance, $\hat{C}_{l,l'}$, are computed from the posterior pdf, $q^{(p-1)}(\mathbf{w}|\mathbf{y})$, in Eq. (25) as

$$\hat{w}_l = \mathsf{E}_{q^{(p-1)}(\mathbf{w}|\mathbf{y})}[w_l], \quad (35)$$

$$\hat{C}_{l,l'} = \mathsf{E}_{q^{(p-1)}(\mathbf{w}|\mathbf{y})}[w_l w_{l'}^H] - \hat{w}_l \hat{w}_{l'}^H, \quad l' \in \hat{\mathcal{S}}. \quad (36)$$

For an efficient iterative optimization, $\mathsf{E}_{q(\theta_l|\mathbf{y})}[\mathbf{a}(\theta_l)]$ [Eq. (34)] is calculated in closed form. A mixture of VMs can approximate the marginal pdfs, $q(\theta_l|\mathbf{y})$, accurately and enable a closed-form calculation (Ref. 46, Sec. IV), as reviewed below. We use $\boldsymbol{\eta}_l = [\eta_{0,l} ... \eta_{M-1,l}]^T$. From Eq. (32), we obtain, cf. Eq. (4),

$$q^{(p)}(\theta_l|\mathbf{y}) \propto f_{\mathrm{VM}}(\theta_l; \eta_{a,l}) \prod_{m=0}^{M-1} \exp\left\{\mathrm{Re}\left(\eta_{m,l}^H e^{jm\theta_l}\right)\right\}$$

$$\propto f_{\mathrm{VM}}(\theta_l; \eta_{a,l}) \prod_{m=1}^{M-1} f_{\mathrm{VM}}(m\theta_l; \eta_{m,l}), \quad (37)$$

where we introduced the parameter $\eta_{a,l}$ of the prior VM pdf. The predicted VM based on the posterior pdfs of the previous time [Eq. (62)] is used as the prior VM. The $f_{\mathrm{VM}}(m\theta_l; \eta_{m,l})$ is referred to as $m$-fold wrapped VM [Eq. (37)] with parameter $\eta_{m,l} = \kappa_{m,l} e^{j\mu_{m,l}}$. The factor corresponding to $m = 0$ is a constant and, thus, omitted in Eq. (37).

An $m$-fold wrapped VM has $m$ modes and is approximated accurately by a mixture of $m$ VM [Ref. 46, Eq. (30)], i.e.,

$$f_{\mathrm{VM}}(m\theta_l; \eta_{m,l}) \approx \sum_{r=1}^{m} \frac{1}{m} f_{\mathrm{VM}}(\theta_l; \tilde{\eta}_{(m,r),l}), \quad (38)$$

where $\tilde{\eta}_{(m,r),l} = \tilde{\kappa}_{m,l} e^{j\tilde{\mu}_{(m,r),l}}$. The $(m, r)$ component corresponds to the mode $r$ of the $m$-fold wrapped VM. The $m$ components of the mixture have equal concentrations but different directions. Matching the characteristic function of the wrapped VM and the mixture of VMs, the single concentration $\tilde{\kappa}_{m,l}$ is the solution of [Ref. 46, Eq. (31)]

$$\frac{I_m(\tilde{\kappa}_{m,l})}{I_0(\tilde{\kappa}_{m,l})} = \frac{I_1(\kappa_{m,l})}{I_0(\kappa_{m,l})}, \quad (39)$$

and the means, $\tilde{\mu}_{(m,r),l}$, $r = 1, ..., m$, are given by

$$\tilde{\mu}_{(m,r),l} = \frac{\mu_{m,l} + 2\pi(r-1)}{m}. \quad (40)$$

Plugging Eq. (38) into Eq. (37) and using the fact that the VM is closed under multiplication [Eq. (5)], the approximate marginal pdf, $q^{(p)}(\theta_l|\mathbf{y})$, reads [Ref. 46, Eq. (33)]

$$q^{(p)}(\theta_l|\mathbf{y}) = \frac{1}{Z_\theta} \sum_{\mathbf{r} \in \mathcal{R}} \exp\left\{\mathrm{Re}\left(\xi_{\mathbf{r},l}^H e^{j\theta_l}\right)\right\}, \quad (41)$$

$$Z_\theta = 2\pi \sum_{\mathbf{r} \in \mathcal{R}} I_0(|\xi_{\mathbf{r},l}|), \quad (42)$$

$$\xi_{\mathbf{r},l} = \eta_{a,l} + \tilde{\eta}_{(1,r_1),l} + \cdots + \tilde{\eta}_{(M-1,r_{M-1}),l}, \quad (43)$$

where indices, $\mathbf{r}$, are from the set $\mathcal{R}$, $\mathbf{r} = (r_1, ..., r_{M-1}) \in \mathcal{R} \triangleq \{1\} \times \{1, 2\} \times \cdots \times \{1, ..., M-1\}$. For example, for $M=4$, we have $\mathcal{R} = \{(1,1,1),(1,1,2),(1,1,3),(1,2,1),(1,2,2),(1,2,3)\}$. When no prior information on $\theta$ is available, e.g., $t=1$, we set $\eta_{a,l} = 0$.

Note that $q^{(p)}(\theta_l|\mathbf{y})$ [Eq. (41)] has the form of a mixture of VMs,

$$q^{(p)}(\theta_l|\mathbf{y}) = \sum_{\mathbf{r}\in\mathcal{R}} C_{\mathbf{r}} f_{\text{VM}}(\theta_l; \xi_{\mathbf{r},l}), \tag{44}$$

where $C_{\mathbf{r}} = 2\pi I_0(|\xi_{\mathbf{r},l}|)/Z_\theta$. Performing the variational Bayes optimization with all $|\mathcal{R}|$ components is intractable. Thus, a suboptimum pruning stage that keeps $D \ll |\mathcal{R}|$ most dominant components is employed,

$$q^{(p)}(\theta_l|\mathbf{y}) = \sum_{d=1}^{D} \tilde{C}_d f_{\text{VM}}(\theta_l; \xi_{d,l}), \tag{45}$$

where $\tilde{C}_d = 2\pi I_0(|\xi_{d,l}|)/\tilde{Z}_\theta$ and $\tilde{Z}_\theta = 2\pi \sum_{d=1}^{D} I_0(|\xi_{d,l}|)$, which gives $\hat{\mathbf{a}}_l$ and corresponding $\hat{\theta}_l$ (Ref. 46, heuristic 1).

To reduce computational complexity, we approximate the mixture of $D$ VMs [Eq. (45)] by a single VM, i.e.,

$$q^{(p)}(\theta_l|\mathbf{y}) \approx f_{\text{VM}}(\theta_l; \tilde{\eta}_l), \tag{46}$$

where $\tilde{\eta}_l = \tilde{\kappa}_l e^{j\tilde{\mu}_l}$. It is assumed that the $D$ VM components are well concentrated and, thus, accurately approximated by a Gaussian [Ref. 64, Eq. (3.5.22)] with mean, $\mu_{d,l} = \arg\{\xi_{d,l}\}$, and variance, $\sigma_{d,l}^2 = 1/|\xi_{d,l}|$. Based on moment matching, we obtain the mean, $\tilde{\mu}_l$, and variance, $\tilde{\sigma}_l^2$, of a single Gaussian,

$$\tilde{\mu}_l = \sum_{d=1}^{D} \tilde{C}_d \mu_{d,l}, \tag{47}$$

$$\tilde{\sigma}_l^2 = \sum_{d=1}^{D} \tilde{C}_d (\sigma_{d,l}^2 + \mu_{d,l}^2) - \tilde{\mu}_l^2. \tag{48}$$

The parameter $\tilde{\eta}_l = 1/\tilde{\sigma}_l^2 e^{j\tilde{\mu}_l}$ fully specifies Eq. (46).

### C. Computing $q^{(p)}(\mathbf{w},\mathbf{s}|\mathbf{y})$

Next, we review calculation of the approximate marginal pdf, $q^{(p)}(\mathbf{w},\mathbf{s}|\mathbf{y})$, in Eq. (29) from approximate marginal pdfs, $q^{(p-1)}(\theta_l|\mathbf{y})$, $l=1,\ldots,L$. Let us introduce the matrix $\mathbf{J}$ and vector $\mathbf{h}$ such that

$$J_{l,l'} = \begin{cases} L, & l=l', \\ \hat{\mathbf{a}}_l^H \hat{\mathbf{a}}_{l'}, & l\neq l', \end{cases} \quad l,l' \in \{1,\ldots,L\}, \tag{49}$$

$$\mathbf{h} = [\hat{\mathbf{a}}_1^H \mathbf{y}, \ldots, \hat{\mathbf{a}}_L^H \mathbf{y}]^T. \tag{50}$$

For each $l=1,\ldots,L$, $\hat{\mathbf{a}}_l$ is calculated based on $q^{(p-1)}(\theta_l|\mathbf{y})$; see Eq. (34). Let $\mathbf{h}_{\mathcal{S}}$ and $\mathbf{J}_{\mathcal{S}}$ be the subvector and submatrix of $\mathbf{h}$ and $\mathbf{J}$, respectively, which only consist of elements corresponding to active indices in $\mathbf{s}$. The approximate marginal pdf, $q^{(p)}(\mathbf{w},\mathbf{s}|\mathbf{y})$, in Eq. (30) is expressed as [Ref. 46, Sec. III B; Ref. 47, Eq. (22)]

$$q^{(p)}(\mathbf{w},\mathbf{s}|\mathbf{y}) \propto \exp\left\{-(\mathbf{w}_{\mathcal{S}} - \hat{\mathbf{w}}_{\mathcal{S}})^H \hat{\mathbf{C}}_{\mathcal{S}}^{-1} (\mathbf{w}_{\mathcal{S}} - \hat{\mathbf{w}}_{\mathcal{S}})\right\}, \tag{51}$$

$$\hat{\mathbf{w}}_{\mathcal{S}} = \nu^{-1} \hat{\mathbf{C}}_{\mathcal{S}} \mathbf{h}_{\mathcal{S}}, \tag{52}$$

$$\hat{\mathbf{C}}_{\mathcal{S}} = \nu \left(\mathbf{J}_{\mathcal{S}} + \frac{\nu}{\tau} \mathbf{I}_{|\mathcal{S}|}\right)^{-1}. \tag{53}$$

Using $q^{(p)}(\mathbf{w},\mathbf{s}|\mathbf{y})$ in Eq. (51) evaluated at the single vector $\hat{\mathbf{s}} \in \mathcal{B}^L$, we obtain the approximate marginal pdf,

$$q^{(p)}(\mathbf{w}|\mathbf{y}) \propto q^{(p)}(\mathbf{w},\hat{\mathbf{s}}|\mathbf{y}). \tag{54}$$

Using Eq. (51) in Eq. (54), the final expression for the approximate marginal pdf, $q^{(p)}(\mathbf{w}|\mathbf{y})$, becomes

$$q^{(p)}(\mathbf{w}|\mathbf{y}) = f_{\text{CN}}(\mathbf{w}_{\hat{\mathcal{S}}}; \hat{\mathbf{w}}_{\hat{\mathcal{S}}}, \hat{\mathbf{C}}_{\hat{\mathcal{S}}}) \prod_{l\notin\hat{\mathcal{S}}} \delta(w_l), \tag{55}$$

which is used for calculating $q^{(p)}(\theta_l|\mathbf{y})$; see Eqs. (35) and (36).

Next, we discuss the calculation of the "best" activation $\hat{\mathbf{s}}$ that defines $\hat{\mathcal{S}}$ on the right-hand side of Eq. (55). For the calculation of this activation, $\hat{\mathbf{s}}$, the ELBO objective function is introduced as

$$O(\mathbf{s}) \triangleq \mathcal{L}\left[q^{(p-1)}(\boldsymbol{\theta},\mathbf{w},\mathbf{s}|\mathbf{y})\right]$$
$$= \mathsf{E}_{q^{(p-1)}(\boldsymbol{\theta},\mathbf{w},\mathbf{s}|\mathbf{y})} \left[\ln \frac{f(\mathbf{y},\boldsymbol{\theta},\mathbf{w},\mathbf{s})}{q^{(p-1)}(\boldsymbol{\theta},\mathbf{w},\mathbf{s}|\mathbf{y})}\right], \tag{56}$$

where $q^{(p-1)}(\boldsymbol{\theta},\mathbf{w},\mathbf{s}|\mathbf{y})$ is from Eq. (24), replacing $q(\theta_l|\mathbf{y})$ and $q(\mathbf{w}|\mathbf{y},\mathbf{s})$ by $q^{(p-1)}(\theta_l|\mathbf{y})$ and $q^{(p)}(\mathbf{w}|\mathbf{y},\mathbf{s})$, respectively. The general expression of $O(\mathbf{s})$ over iteration $p$ is given in Ref. 46, Eq. (21). The activation $\hat{\mathbf{s}}$ at iteration $p$ is the maximizer of $O(\mathbf{s}) = \mathcal{L}[q^{(p-1)}(\boldsymbol{\theta},\mathbf{w},\mathbf{s}|\mathbf{y})]$, i.e.,

$$\hat{\mathbf{s}} = \operatorname{argmax}_{\mathbf{s}} O(\mathbf{s}). \tag{57}$$

Finding the global maximum of $O(\mathbf{s})$ implies an exhaustive search over all $2^L$ possible values of $\mathbf{s} \in \mathcal{B}^L$ and is infeasible. Alternatively, a local maximum is found by employing an iterative search. This search starts from the solution at iteration $p-1$ and iteratively flips single elements greedily until no further maximization is possible (Ref. 46, Appendix A; Ref. 47, Appendix A 2).

### D. Estimating model parameters

The ELBO, $\mathcal{L}[q^{(p)}(\boldsymbol{\theta},\mathbf{w},\mathbf{s}|\mathbf{y})]$ [Eq. (22)], is a function of the model parameters, $\nu$ and $\tau$, thus, $\mathcal{L}^{(p)}[\nu,\tau] \triangleq \mathcal{L}[q^{(p)}(\boldsymbol{\theta},\mathbf{w},\mathbf{s}|\mathbf{y})]$. Following Ref. 46, Sec. III C, $\nu$ and $\tau$ are estimated by maximizing $\mathcal{L}^{(p)}[\nu,\tau]$ at the end of each step $p$. This maximization is performed over each parameter by setting $\partial\mathcal{L}^{(p)}[\nu,\tau]/\partial\nu = 0$ and $\partial\mathcal{L}^{(p)}[\nu,\tau]/\partial\tau = 0$, whereby

$$\hat{\nu} = \frac{1}{M}\left\|\mathbf{y} - \sum_{l\in\hat{\mathcal{S}}} \hat{w}_l \hat{\mathbf{a}}_l \right\|_2^2 + \frac{\operatorname{tr}\left[\mathbf{J}_{\hat{\mathcal{S}}} \hat{\mathbf{C}}_{\hat{\mathcal{S}}}\right]}{M} + \sum_{l\in\hat{\mathcal{S}}} |\hat{w}_l|^2 \left(1 - \frac{\|\hat{\mathbf{a}}_l\|_2^2}{M}\right), \tag{58}$$

$$\hat{\tau} = (\hat{\mathbf{w}}_{\hat{\mathcal{S}}}^H \hat{\mathbf{w}}_{\hat{\mathcal{S}}} + \operatorname{tr}(\hat{\mathbf{C}}_{\hat{\mathcal{S}}}))/|\hat{\mathcal{S}}|. \tag{59}$$

## V. SEQUENTIAL VARIATIONAL BAYESIAN ESTIMATION

In this section, we present a joint message passing method that combines BP and MF message passing approaches for sequential variational line spectral estimation.

### A. Prediction

Let us introduce the sequence of all of the measurements up to time $t-1$ as $\mathbf{y}_- \triangleq (\mathbf{y}_1, ..., \mathbf{y}_{t-1})$, $\hat{s}_{t-1} \in \mathcal{B}^L$ as the set of activation indices and $q(\theta_{l,t-1}|\mathbf{y}_-)$, $l \in \hat{\mathcal{S}}_{t-1}$ as the approximate marginal pdfs at $t-1$. From $\hat{s}_{t-1}$, we directly get the approximate marginals pmfs, $q(s_{l,t-1}|\mathbf{y}_-) = \delta(s_{l,t-1} - \hat{s}_{l,t-1})$. At time $t$, the proposed sequential Bayesian estimation method predicts the approximate marginal pdfs $q(\theta_{l,t}|\mathbf{y}_-)$ and $q(s_{l,t}|\mathbf{y}_-)$ from $q(\theta_{l,t-1}|\mathbf{y}_-)$ and $q(s_{l,t-1}|\mathbf{y}_-)$, respectively, in a *prediction step*. We present the sequential processing as an instance of BP message passing. The approach is similar to the sequential Bayesian inference in a state space model, e.g., message passing for the Kalman filter (Ref. 56, Sec. IV C).

Following BP message passing rules,[56] for the pdfs of PAs $\theta_{l,t}$, $l \in \hat{\mathcal{S}}_{t-1}$, we predict $q(\theta_{l,t}|\mathbf{y}_-)$ using the state transition pdf $f(\theta_{l,t}|\theta_{l,t-1})$ and the posterior pdf $q(\theta_{l,t-1}|\mathbf{y}_-)$ [Eq. (46)],

$$q(\theta_{l,t}|\mathbf{y}_-) = \int f(\theta_{l,t}|\theta_{l,t-1}) q(\theta_{l,t-1}|\mathbf{y}_-) d\theta_{l,t-1}$$
$$= \int f_{\mathrm{VM}}(\theta_{l,t}|\theta_{l,t-1}; 0, \kappa_{\mathrm{r}})$$
$$\times f_{\mathrm{VM}}(\theta_{l,t-1}; \hat{\theta}_{l,t-1}, \hat{\kappa}_{l,t-1}) d\theta_{l,t-1}, \quad (60)$$

where $\hat{\eta}_{l,t-1} = \hat{\kappa}_{l,t-1} e^{j\hat{\theta}_{l,t-1}}$ [Eq. (46)]. The state transition is modeled by the VM transition pdf $f(\theta_{l,t}|\theta_{l,t-1}) = f_{\mathrm{VM}}(\theta_{l,t}|\theta_{l,t-1}; 0, \kappa_{\mathrm{r}})$.

The dynamic DOAs are modeled by a state transition model,

$$\theta_{l,t} = \theta_{l,t-1} + \mathsf{r}_{l,t}, \quad (61)$$

where $\mathsf{r}_{l,t}$ is an iid sequence of von Mises distributed random variables with zero mean and $\kappa_{\mathrm{r}}$. For example, $\kappa_{\mathrm{r}} = 148$ provides standard deviation of $1.499°$ for a half-wavelength spacing, i.e., $(\sigma_{\mathrm{r}} = 0.0822\,\mathrm{rad};\ \sin^{-1}\sigma_{\mathrm{r}}c/\omega d = 1.499°)$. From the state transition model [Eq. (61)], we obtain the transition pdf $f(\theta_{l,t}|\theta_{l,t-1}) = f_{\mathrm{VM}}(\theta_{l,t}|\theta_{l,t-1}; 0, \kappa_{\mathrm{r}})$ [Eq. (60)].

The VMs $f_{\mathrm{VM}}(\theta_{l,t}|\theta_{l,t-1})$ and $f_{\mathrm{VM}}(\theta_{l,t-1})$ typically have large $\kappa$, which is analogous to Gaussians having a small variance $\sigma^2 = 1/\kappa$, i.e., these VMs are concentrated about their mean. An approximation of Eq. (60) is obtained as [cf. Eq. (46)]

$$q(\theta_{l,t}|\mathbf{y}_-) \approx f_{\mathrm{VM}}(\theta_{l,t}; \hat{\eta}'_{l,t}), \quad (62)$$

where $\hat{\eta}'_{l,t}$ is obtained by approximating $f_{\mathrm{VM}}(\theta_{l,t}|\theta_{l,t-1})$ by a zero-mean Gaussian with variance $1/\kappa_{\mathrm{r}}$ and $f_{\mathrm{VM}}(\theta_{l,t-1}; \hat{\eta}_{l,t-1})$ by a $\hat{\theta}_{l,t-1}$-mean Gaussian with variance $1/\hat{\kappa}_{l,t-1}$ [Eq.

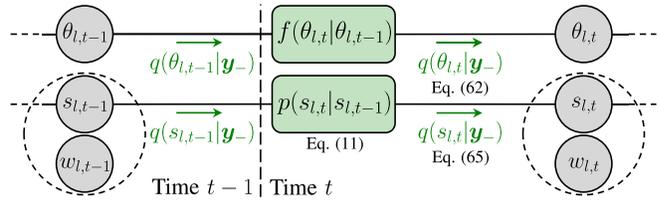

FIG. 3. (Color online) BP procedure for sequential processing, $l \in \{1, ..., L\}$. A portion of the factor graph in Fig. 1 is shown.

(60)]. Performing closed-form integration gives a $\hat{\theta}_{l,t-1}$-mean Gaussian with variance $1/\kappa_{\mathrm{r}} + 1/\hat{\kappa}_{l,t-1}$ using Eq. (61). Approximating the Gaussian by a VM, $\hat{\eta}'_{l,t}$ becomes

$$\hat{\eta}'_{l,t} = (1/\kappa_{\mathrm{r}} + 1/\hat{\kappa}_{l,t-1})^{-1} e^{j\hat{\theta}_{l,t-1}}. \quad (63)$$

For prediction of $\mathsf{s}_{l,t}$, we compute

$$q(s_{l,t}|\mathbf{y}_-) = \sum_{s_{l,t-1} \in \{0,1\}} p(s_{l,t}|s_{l,t-1}) q(s_{l,t-1}|\mathbf{y}_-). \quad (64)$$

Using the transition pmf $p(s_{l,t}|s_{l,t-1})$ [Eq. (11)] and exploiting $q(s_{l,t-1}|\mathbf{y}_-) = \delta(s_{l,t-1} - \hat{s}_{l,t-1})$, Eq. (64) becomes

$$q(s_{l,t}|\mathbf{y}_-) = \begin{cases} p^d \hat{s}_{l,t-1} + (1-p^a)(1-\hat{s}_{l,t-1}), & s_{l,t} = 0, \\ p^a(1-\hat{s}_{l,t-1}) + (1-p^d)\hat{s}_{l,t-1}, & s_{l,t} = 1. \end{cases}$$
(65)

At time $t=1$, $q(\theta_{l,t-1}|\mathbf{y}_-)$ in Eq. (60) and $q(s_{l,t-1}|\mathbf{y}_-)$ in Eq. (64) are replaced by the initial priors $f_{\mathrm{VM}}(\theta_{l,0})$ and $p(s_{l,0})$, respectively. Note that according to Eq. (65), we have $\rho_{l,t} = p^a$ if $l \in \{1,...,|\hat{\mathcal{S}}_{t-1}|\}$ and $1-p^d$ otherwise, cf. Eq. (6). At time $t=1$, we set $\rho_{1,t} = \cdots = \rho_{L,t} = \rho$.

The approximate marginal pdfs and pmfs can be interpreted as BP messages passed along the edges of the factor graphs in Figs. 1 and 3, which appear in green.

### B. Update

The prediction step is followed by an *update step* that computes $\hat{s}_t \in \mathcal{B}^L$ and $q(\theta_{l,t}|\mathbf{y})$, $l \in \hat{\mathcal{S}}_t$. Iterative MF message passing, according to Eqs. (30) and (27), is performed. The update step calculates the approximate marginal pdfs, $q(\theta_{l,t}|\mathbf{y})$, and pmfs, $q(s_{l,t}|\mathbf{y})$, in closed form; see Secs. IV B and IV C.

The iterative MF message passing in Sec. IV needs an initial guess for $q(\theta_{l,t}|\mathbf{y})$ [Eq. (27)] for iteration $p=0$ with this initialization; see Sec. VI. The predicted pdfs, $q(\theta_{l,t}|\mathbf{y}_-)$, are incorporated for MF message passing for the update step and used as the initialization ($p=0$) for the following iteration, i.e., $q^{(0)}(\theta_{l,t}|\mathbf{y}) = q(\theta_{l,t}|\mathbf{y}_-)$ [Eq. (27)].

Next, the predicted pdfs, $q(\theta_{l,t}|\mathbf{y}_-)$, are used for computing $q^{(p)}(\theta_l|\mathbf{y})$ in Sec. IV B as the prior VM pdfs. We introduced the parameter $\eta_{a,l}$ of the prior VM pdf in Eqs. (37) and (41). $q(\theta_{l,t}|\mathbf{y}_-)$ is approximated by a VM [Eq. (62)] with $\hat{\eta}'_{l,t}$ [Eq. (63)]. For $q^{(p)}(\theta_l|\mathbf{y})$ [Eq. (41)], the prior VM pdfs are used with $\eta_{a,l} = \hat{\eta}'_{l,t}$.

ALGORITHM 1. SVALSE.

---

**Input:** Measurements $y_t$,
    *Prior information* [$t > 1$, *prediction* step (Sec. V A)]
    $q(\theta_{l,t}|y_-)$ [Eqs. (62) and (63)] using $\hat{a}_{l,t-1}$, $l \in \{1, ..., L'\}$,
    $L'$ is the number of DOA transitions (Sec. VI)
    $p(s_t|y_-)$ [Eq. (65)] using $p^a$ and $p^d$
**Output:** Number of sources estimates $|\hat{\mathcal{S}}|$, DOA estimates $\hat{\theta}_{\hat{\mathcal{S}}}$, complex amplitude estimates $\hat{w}_{\hat{\mathcal{S}}}$ at time $t$

1:   *Incorporate the predicted information* ($t > 1$)
    $\hat{a}_l$ from time $t - 1$, $l \in \{1, ..., L'\}$
    $\rho_l = p^a$ if $l \in \{1, ..., |\hat{\mathcal{S}}_{t-1}|\}$ and $1 - p^d$ otherwise.
2:   Initialize $\{\hat{a}_l\}_{l=L'+1}^L$, $\nu$, and $\tau$ (Sec. VI)
3:   Compute $J$ [Eq. (49)] and $h$ [Eq. (50)] for $q(w, s|y)$ (Sec. IV C)
4:   **repeat**
5:     [*Update* step (Sec. V B)]
      Update $s$ [Eq. (57)], followed by $w_\mathcal{S}$ [Eq. (52)] and $\hat{C}_\mathcal{S}$ [Eq. (53)]
6:     Update $\nu$ [Eq. (58)] and $\tau$ [Eq. (59)] (Sec. IV D)
7:     For all $l \in \mathcal{S}$, update $q(\theta_l|y) = f_{VM}(\theta_l; \hat{\eta}_l)$ [Eq. (46)]
      (*Incorporate the predicted information* ($t > 1$) for the prior VM pdfs, $\eta_{a,l} = \hat{\eta}'_{l,t}$ [Eqs. (41) and (63)] (Sec. IV B)
8:   **until** stopping criterion
9:   **return** $|\hat{\mathcal{S}}|$, $\hat{\theta}_{\hat{\mathcal{S}}}$, and $\hat{w}_{\hat{\mathcal{S}}}$

---

The MF message passing approaches updates $q(\theta_{l,t-1}|y_-)$ and $q(s_{l,t-1}|y_-)$ for DOA estimation at time $t - 1$. The BP approach is used for the prediction using $q(\theta_{l,t}|y_-)$ [Eq. (60)] and $q(s_{l,t}|y_-)$ [Eq. (65)], $l \in \hat{\mathcal{S}}_{t-1}$, as passing messages. The MF message passing updates $q(\theta_{l,t}|y)$ [Eq. (27)] and $q(s_{l,t}|y)$ [Eq. (30)].

The resulting combined BP and MF message passing approach calculates accurate approximations of closed-form marginal pdfs. It can be seen as an instance of the general framework for combining BP with MF message passing.[63]

The procedure for iteratively updating the factors of $\{\theta, w, s\}$ and estimating $\{\nu, \tau\}$ is summarized in Algorithm 1.

## VI. IMPLEMENTATION ASPECTS

Before the iteration starts for variational Bayesian estimation in Sec. IV, at iteration $p = 0$, (S)VALSE requires initial guesses, including $\{\nu, \tau, \{q^{(0)}(\theta_l|y)\}_{l=1}^L, K\}$. Regarding the initialization, SVALSE uses the same scheme as was used for VALSE.[46]

### A. Initializing the von Mises pdf

Regarding the initial guess for $\{q(\theta_l|y)\}_{l=1}^L$ at initialization ($p = 0$) without knowledge of $w$, we assume an improper flat prior for $w$ [Ref. 46, Eq. (38)], i.e.,

$$q(\theta_l|y) = \int f_{CN}(y; wa(\theta_l), \nu) dw \propto \exp\left(\frac{|y^H a(\theta_l)|^2}{\nu N}\right). \quad (66)$$

Similarly for the VM approximation [Eq. (32)] in Sec. IV B, let $\mathcal{I}' = \{l_i - l_j | l_i, l_j \in \mathcal{I}, l_i > l_j\}$ with cardinality $L' = L - 1$ and the vector-valued function, $a' : [-\pi, \pi) \to \mathbb{C}^{L'}$, $\theta \to a'(\theta) \triangleq (e^{j\theta l}|l \in \mathcal{I}')^T$. Then, Eq. (66) becomes

$$q(\theta_l|y) \propto \exp\left(\text{Re}\left\{\frac{2}{\nu}\gamma^H a'(\theta_l)\right\}\right), \quad (67)$$

where for each $l = 1, ..., L'$, $\gamma_l = (1/L)\sum_{(l_i,l_j)\in\mathcal{I}_l} y_{l_i} y_{l_j}^H$ with $\mathcal{I}_l = \{(l_i, l_j)|1 \leq l_i, l_j \leq L, l_i - l_j = l\}$. Using $\eta = (2/\nu)\gamma$ in Eq. (32) from Eq. (67), we obtain $q(\theta_1|y)$ and $\hat{a}_1(= \mathbb{E}_{q(\theta_1|y)}[a(\theta_1)])$ [Eq. (46)] and estimate $\hat{w}_1$ based on Eqs. (52) and (53) using $J$ [Eq. (49)] and $h$ [Eq. (50)]. Next, we update $q(\theta_2|y)$ based on the residual $z_1 = y - \hat{w}_1\hat{a}_1$. For the $l$th component, we update $q(\theta_l|y)$ based on the residual $z_{l-1} = y - \sum_{l'=1}^{l-1} \hat{w}_{l'}\hat{a}_{l'}$,

$$q(\theta_l|y) \propto \exp\left(\frac{|z_{l-1}^H a(\theta_l)|^2}{\nu N}\right). \quad (68)$$

To obtain $q(\theta_l|y)$ and $\hat{a}_l$, we use Ref. 46, heuristic 2.

### B. Initializing the noise variance, $\nu$

The noise variance, $\nu$, influences the sparsity of the solution. The solution becomes sparse for high $\nu$ as it admits more noise and fits the data with fewer components. High initialized $\nu$ excludes many potential DOAs and results in missing DOAs. Low initialized $\nu$ considers more DOAs and needs more computations for convergence.

The original VALSE[46] builds a Toeplitz estimate of $\mathsf{E}[yy^H]$ by averaging the diagonal elements of $yy^H$. Then, $\nu$ is initialized with the average of the lower quarter of the Toeplitz matrix eigenvalues. This calculation provides random performance depending on the eigenvalues, e.g., it can initialize $\nu$ too high, which will exclude all potential DOAs.

At initialization ($p = 0$) for $\nu$, we assume a measurement-to-noise ratio of 20 dB, i.e., $10 \log_{10}[\|y\|_2^2/M\nu] = 20$ or $\nu = \|y_2^2\|/M/10^{20/10}$. At later iterations, $\nu$ is estimated from Eq. (58).

The variance, $\tau$, of the Gaussian distributed amplitudes, $w$ [Eq. (6)], is initialized with $\tau = (y^Hy/M - \nu)/(\rho L)$, obtained from $y^Hy/M = \rho L\tau + \nu$ [Eq. (2)]. The activation probability, $\rho$, of the Bernoulli variable, $s$, is initialized as $\rho = 0.5$ at time $t = 1$.

### C. Information transfer between time steps

DOAs are estimated using $L$ PAs, and the sequential method propagates the PAs through the prediction and update steps. At time $t - 1$, after update, we obtain $L$ PAs with $\{q(\theta_l|y)\}_{l=1}^L$, among which nonzero entries in $\hat{s}_{t-1} \in \mathcal{B}^L$ are active and their DOAs are estimated. The PAs from time $t - 1$ are transferred to the prediction at time $t$ and used for the update at iteration $p = 0$. For the sequential process, we propagate only the active PAs.

At time $t - 1$, during iterations $p \in \{1, ..., P\}$, we record the activations $s_{t-1}^{(p)}$. Only active components and components with the activation history ($s_{l,t-1}^{(p)} = 1$) are propagated to the subsequent time step in form of the corresponding VM $q(\theta_{l,t-1}|y)$. The rest of the components out of $L$ are not propagated. At time $t$, these are re-initialized using the VM initialization scheme[68] based on the propagated VM. The activation probability is propagated based on Eq. (65).

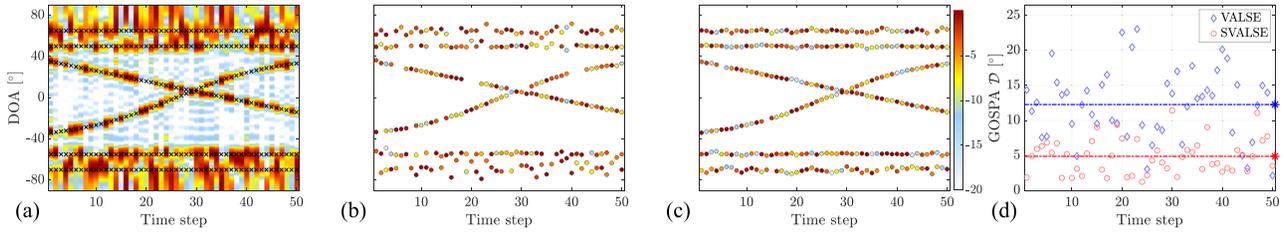

FIG. 4. (Color online) DOA versus time for (a) true ($\times$) and CBF (background), (b) nonsequential VALSE (Refs. 46 and 47), and (c) proposed SVALSE in the simulated scenario. (d) The GOSPA errors [Eq. (70)] are shown for each individual time step and averaged over the time step data for nonsequential VALSE, 12.27° and SVALSE, 4.87°.

This information transfer allows the appearance of newly detected sources at each time step. The dynamic DOA performance is evaluated next.

## VII. SIMULATION

### A. Setup

We considered $L = 15$ sensors and $K = 15$ potential DOAs. The sensors form a uniform linear array with a sensor spacing of 3.75 m, half-wavelength spacing, and sound speed $c = 1500$ m/s. In Figs. 4–6, we simulated six DOAs ($K_t = 6$ for all $t$) with initial angles $\boldsymbol{\beta}_0^\star = [-70°\ -55°\ -40°\ 35°\ 50°\ 65°]^T$ that transmit at 200 Hz. The two "inner" DOAs $k \in \{3, 4\}$ with initial positions $\beta_{3,0}^\star = -40°$ and $\beta_{4,0}^\star = 35°$ are dynamic. Sources can suddenly appear and disappear. A scenario in which two out of six DOAs are deactivated and a scenario in which the first two, second two, and third two DOAs from 90° are deactivated sequentially is simulated in Figs. 5 and 6. The variance of source amplitudes is fixed to $\tau = 1$. We also simulated three static DOAs located at $\boldsymbol{\beta}_t^\star = [-3°\ 2°\ 60°]^T$ with deterministic amplitude of ten; see Figs. 8 and 10(b). The noise variance, $\nu$, is set to obtain a signal-to-noise ratio (SNR),

$$\text{SNR} = 20 \log_{10}\left[\left\|\sum_{l=1}^{K_t} w_{l,t}\boldsymbol{a}(\theta_{l,t})\right\|_2 \bigg/ \|\boldsymbol{u}_t\|_2\right]. \quad (69)$$

### B. Metrics for DOA performance evaluation

We evaluate and compare the DOA performance based on the root mean square error (RMSE) and the generalized optimum sub-pattern assignment metric (GOSPA).[70]

GOSPA assesses the number of estimated DOAs (cardinality) and the estimation error in angle. We can encounter when the true and estimated DOAs have different cardinalities, $K \neq \hat{K}$, and GOSPA considers false and missed DOAs. False DOAs are the estimated DOAs but with an estimation error above $c/2$ ($c$ is a user-chosen constant, here, 10°), and missed DOAs are DOAs not estimated. GOSPA sets the cardinality mismatch cost for false and missed DOAs and assigns the estimated DOAs to the true DOAs with the minimum measure.

Let $\vartheta_a \in \{1, ..., K\} \times \{1, ..., \hat{K}\}$ be assignment set between two sets with $K$ and $\hat{K}$ elements, respectively, which has the property that $(i, j), (i, j') \in \vartheta_a$ implies $j = j'$ and $(i, j), (i', j) \in \vartheta_a$ implies $i = i'$.

GOSPA is a sum of errors for the DOAs, assigned to the true DOAs, and a penalty for missed and false DOAs. The GOSPA $\mathcal{D}[\vartheta_\text{true}, \vartheta_\text{est}]$ for true DOAs $\vartheta_\text{true} = \{\theta_k\}_{k=1}^K$ and estimated DOAs $\vartheta_\text{est} = \{\hat{\theta}_k\}_{j=1}^{\hat{K}}$ is calculated [by setting the parameters $\alpha = 2$ and $p = 1$ in Ref. 70, Eq. (1)],

$$\mathcal{D}[\vartheta_\text{true}, \vartheta_\text{est}] = \mathcal{D}_\text{dist} + \mathcal{D}_\text{miss} + \mathcal{D}_\text{false}$$
$$= \min_{\vartheta_a \in \Theta_{|\vartheta_\text{true}||\vartheta_\text{est}|}} \left[\sum_{(i,j)\in\vartheta_a} d(\theta_i, \hat{\theta}_j)\right.$$
$$\left. + \frac{c}{2}(|\vartheta_\text{true}| + |\vartheta_\text{est}| - 2|\vartheta_a|)\right], \quad (70)$$

$$\mathcal{D}_\text{miss} = c(|\vartheta_\text{true}| - |\vartheta_a|)/2, \quad (71)$$

$$\mathcal{D}_\text{false} = c(|\vartheta_\text{est}| - |\vartheta_a|)/2, \quad (72)$$

where $\Theta_{|\vartheta_\text{true}||\vartheta_\text{est}|}$ is the set of all possible combinations of indices. DOAs that remain unassigned are either missed, $\mathcal{D}_\text{miss}$, or false DOAs, $\mathcal{D}_\text{false}$, and these are penalized by the cardinality mismatch cost $c/2$. The first term, $\mathcal{D}_\text{dist}$ [Eq. (70)], represents a measure of the distances (differences) between assigned and true DOAs, which is an arbitrary "inner metric" of the distance between $\theta_i$ and $\hat{\theta}_j$. Here, we use the absolute errors, i.e.,

$$\mathcal{D}_{\text{dist}} = \sum_{(i,j)\in\vartheta_a} |\theta_i - \hat{\theta}_j|. \tag{73}$$

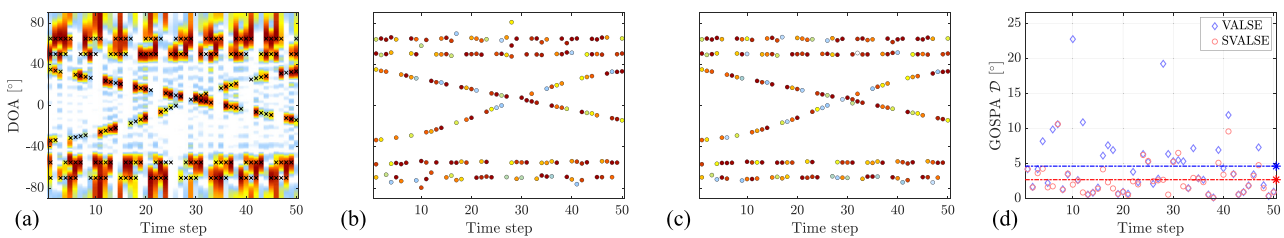

FIG. 5. (Color online) The same as that in Fig. 4 but for each time that two out of six DOAs are deactivated for time $t > 1$. The GOSPA errors averaged over the time step are for nonsequential VALSE, 4.62° and SVALSE, 2.69°.

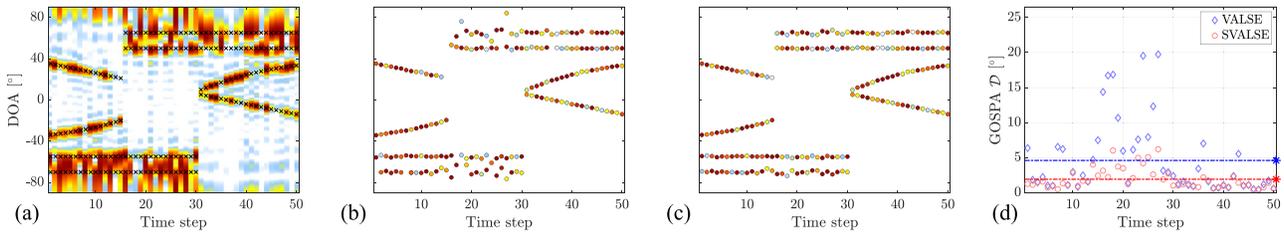

FIG. 6. (Color online) The same as that in Fig. 4 but two DOAs with initial angles [65° 50°], [35° − 40°], and [−55° − 70°] are deactivated sequentially. The GOSPA errors averaged over the time step are for nonsequential VALSE, 4.60° and SVALSE, 1.91°.

We also compare the RMSE, which is calculated as

$$\mathcal{D}_{\text{RMSE}} = \min_{\vartheta_a \in \Theta_{|\vartheta_{\text{true}}||\vartheta_{\text{est}}|}} \left[ \frac{1}{|\vartheta_{\text{true}}|} \left[ \sum_{(i,j)\in\vartheta_a} \left(\theta_i - \hat{\theta}_j\right)^2 + c'^2(|\vartheta_{\text{true}}| - |\vartheta_a|) \right] \right]^{1/2}, \tag{74}$$

where $c'$ is a maximum DOA error for missed or false DOAs that remain unassigned. The DOA assignment is performed using the Hungarian method for optimal point assignment.[71] In Eqs. (70) and (74), we used $c = c' = 10°$.

### C. Algorithms

In the implementation of the VALSE with sequential processing, we set $p^a = 0.10$ and $p^d = 0.25$. For all of the PAs, we consider a state transition function, $f_{\text{VM}}(\theta_{l,t}|\theta_{l,t-1})$ [Eq. (61)], where $\kappa_r = 148$ (this gives $\sigma_r = 0.0822$ rad; $\sin^{-1}(\sigma_r c/\omega d) = 1.499°$). As reference methods, we use SBL,[13] sequential sparse Bayesian learning (SSBL),[25] and VALSE.[46] In SBL and SSBL, we use the steering vectors corresponding to potential DOAs $\theta_{l,t} \in \{−90°, −89.5°, \ldots 89.5°, 90°\}$, $l \in \mathcal{I} \triangleq \{0, \ldots, 360\}$. After performing SBL or SSBL processing, we obtain final DOA estimates by locating local maximum source magnitudes above threshold 1% of the maximum. Nonsequential VALSE is entirely parameter free. Gridless sparse methods[19,20,26,42–45] do not use statistical information for sequential processing, and the VALSE methods[47–49] do not consider time-varying DOAs and, thus, are omitted.

### D. Performance with dynamic DOAs

Figure 4 shows a single simulation of conventional beamforming (CBF), nonsequential VALSE,[46,47] and the proposed SVALSE. The sequential methods can more reliably localize DOAs near to endfire of the array, i.e., at −70°, −55°, 50°, and 65°. In sequential processing, DOA detection is supported by prior information from previous steps. The proposed SVALSE localizes DOAs near the endfire accurately.

Dynamic sources involve sources that suddenly appear and disappear. A scenario in which two out of six DOAs are deactivated and a scenario in which the first two, second two, and third two DOAs from 90° are deactivated sequentially is simulated in Figs. 5 and 6. Although the prediction step in sequential processing uses the previously estimated DOAs, which are suddenly deactivated in the current measurement, the update step incorporates the current measurement and filters the deactivated DOAs.

The proposed SVALSE localizes dynamic DOAs accurately. Appearings in Figs. 4–6(d) are comparisons of VALSE and SVALSE with GOSPA [Eq. (70)] versus time. The error of the proposed SVALSE is lower than VALSE and more stable.

Figure 7 shows the effects of the (de)activation probability, $\{p^d, p^a\}$ [Eq. (11)], on the SVALSE. $\{p^d, p^a\}$ selection controls the level of sparsity. The analytic solution is difficult to answer. However, incorporating prior information ($0 < p^d < 1, 0 < p^a < 1$) improves estimation accuracy for SVALSE over nonsequential VALSE.

The deactivation probability, $p^d$, mitigates DOAs with the activation history; see Figs. 7(a) and 7(b). A small value of $p^d$ causes the DOAs with the activation history to remain. When the true DOAs disappear, the method still estimates the false DOAs; see Fig. 7(a). A large value of $p^d$ removes the DOAs with the activation history and makes the method miss the true DOAs; see Fig. 7(b).

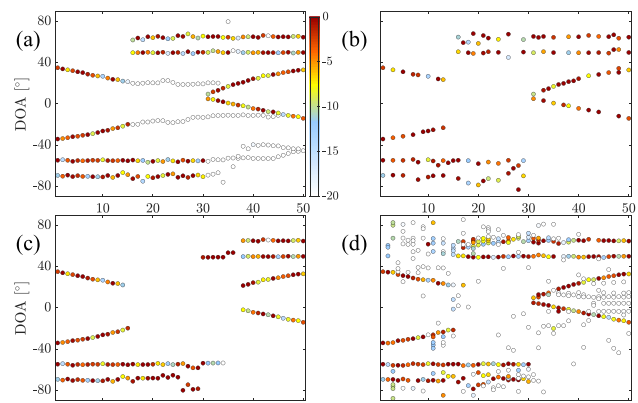

FIG. 7. (Color online) Effect of varying the (de)activation probability $\{p^d, p^a\}$ [Eq. (11)] for SVALSE using the scenario in Fig. 6. GOSPA (RMSE) averaged over the 50 time step data for (a) $\{0, 0.10\}$, 10.74° (0.83°); (b) $\{1, 0.10\}$, 11.32° (6.67°); (c) $\{0.25, 0\}$, 8.06° (5.36°); and (d) $\{0.25, 1\}$, 25.79° (1.60°).

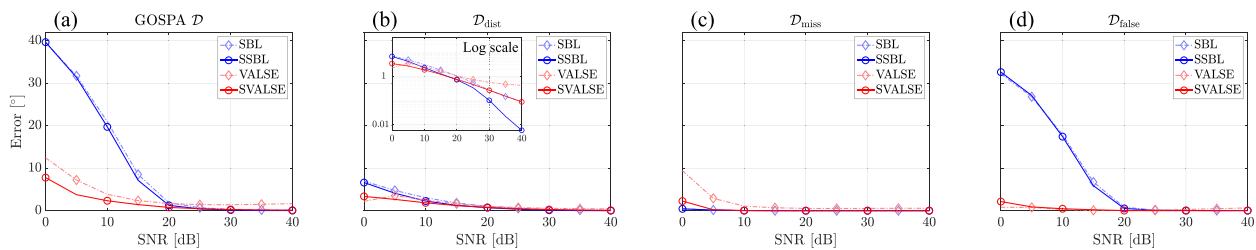

FIG. 8. (Color online) (a) Total GOSPA error and its (b) distance error, (c) missed, and (d) false DOA error contributions versus SNR of simulated methods in scenario 1 are shown.

The activation probability, $p^a$, activates newly updated DOAs; see Figs. 7(c) and 7(d). When DOAs with the activation history miss the true DOAs, the newly updated DOAs act as a buffer. A small value of $p^a$ hinders the method to respond to DOAs newly appearing; see Fig. 7(c). A large value of $p^a$ causes overfitting and makes the method overestimate the model order; see Fig. 7(d).

### E. SNR performance

We show DOA performance results versus SNR based on the GOSPA [Eq. (70)], Figs. 8 and 9, the RMSE [Eq. (74)], and Fig. 10. We set the noise variance, $\nu$, such that SNRs $(0, 5, \ldots, 40)$ dB are obtained. "Scenario 1" in Figs. 8 and 10(a) has three static DOAs located at $\boldsymbol{\beta}_t^\star = [-3^\circ\ 2^\circ\ 60^\circ]^T$, which all have the same deterministic amplitude of ten, and "scenario 2" in Figs. 9 and 10(b) has six DOAs with the variance of amplitudes $\tau = 1$ as in Fig. 4. We considered $t = 50$ time steps as in Fig. 4 for a single run and 100 simulation runs.

Figures 8 and 9 show the mean GOSPA, $\mathcal{D}$ [Eq. (70)], and its distance, $\mathcal{D}_{\mathrm{dist}}$, missed DOAs, $\mathcal{D}_{\mathrm{miss}}$, and false DOAs, $\mathcal{D}_{\mathrm{false}}$, error contributions (averaged over time and simulation runs) versus SNR. It can be seen that (S)VALSE yields lower GOSPA errors than (S)SBL at SNR values below 10 dB. This is because for low SNR values, (S)SBL overestimates the number of DOAs; see Figs. 8(d) and 9(d). SVALSE performs slightly worse than SSBL for high SNR. Grid-based SSBL has an advantage over the proposed gridless method in $\mathcal{D}_{\mathrm{dist}}$ [Figs. 8(b) and 9(b)] because static DOAs in both scenarios are just on the grid without mismatch, resulting in 0° errors.

The GOSPA of (S)VALSE shows low values at SNR below 10 dB [Figs. 8(b) and 9(b)] as $\mathcal{D}_{\mathrm{dist}}$ considers only successfully estimated-and-assigned DOAs for its computation. Further, (S)VALSE underestimates DOAs, and only successfully estimated DOAs go into $\mathcal{D}_{\mathrm{dist}}$, providing low errors. Due to the error contribution of the missed DOAs [Figs. 8(c) and 9(c)], the results have a high GOSPA [Figs. 8(a) and 9(a)].

Figure 10 shows the mean RMSE [Eq. (74); averaged over time and simulation runs] versus SNR. For low SNR values, (S)SBL provides lower RMSE errors than (S)VALSE. This is because for low SNR, (S)SBL overestimates the number of DOAs [Figs. 8(d) and 9(d)], and it is likely to estimate DOAs lower than the maximum DOA error. For high SNR, SVALSE outperforms (S)SBL in Fig. 10(b), as SVALSE is gridless and estimates the number of DOAs accurately.

The proposed SVALSE outperforms all of the other methods compared to (S)SBL and this is due to SVALSE being gridless and accurately estimating the number of DOAs. The proposed sequential processing and initialization enhance the accuracy and estimation of the number of DOAs relative to VALSE.

### VIII. REAL DATA PERFORMANCE

The time-varying DOA performance of the SVALSE is compared with CBF using experimental data; see Figs. 11 and 12. The data are from the SWellEx-96 Event S5 recorded at 23:15–00:30 GMT on 10–11 May 1996, which was performed west of Point Loma, CA.[25,72,73] A vertical uniform linear array (VLA) recorded the acoustic data with $M = 64$ sensors with spacing $d = 1.875$ m and spanning a depth of 94.125–212.25 m (element 43 was corrupted and excluded). A shallow source was towed from 9 km

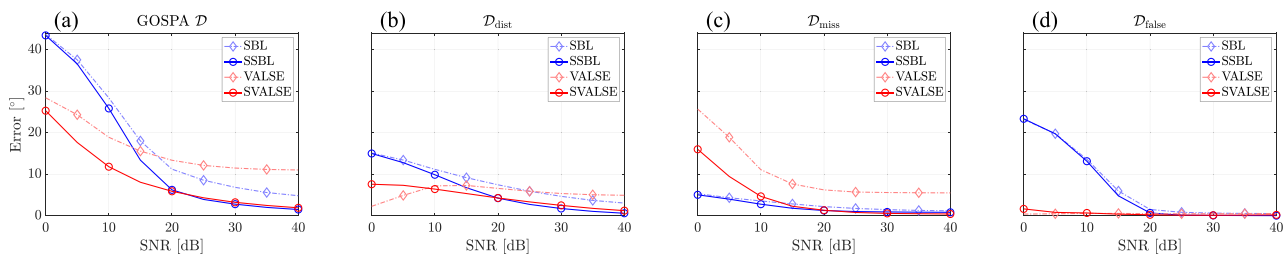

FIG. 9. (Color online) The same as that in Fig. 8 but in scenario 2.

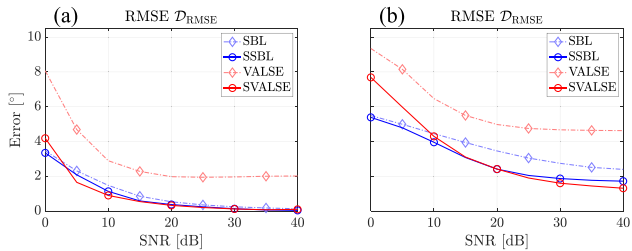
FIG. 10. (Color online) RMSE versus SNR of simulated methods in (a) scenario 1 and (b) scenario 2 are shown.

southwest to 3 km northeast of the array at 2.5 m/s with the closest point of approach at 900 m.

The shallow source transmitted nine tones at frequencies {109, 127, 145, 163, 198, 232, 280, 335, 385} Hz. We focus on the signal component at 198 Hz. The data were sampled at 1500 Hz, and the record at 23:21–00:24 GMT is divided into nonoverlapping 350-time steps; see Fig. 11(c). Each measurement vector is obtained using the discrete Fourier transform with $2^{14}$ samples.

To analyze the time-varying DOA structure, see Fig. 11; we simulate the acoustic field using underwater acoustic propagation models, the Kraken normal mode model, and the Bellhop ray-tracing model.[74] Both methods require environmental information, including sound speed profile, bathymetry, and geo-acoustic parameters, and use the information as in Ref. 75. The characteristics of the acoustic environment, including the water column sound speed profile [Fig. 11(b)], cause waveguide multipaths [Fig. 11(a)]. As the source moves closer to the VLA (increasing time step), the absolute DOA increases. The critical angle for this environment is $\cos^{-1}(1488/1600) = 21.6°$, beyond which multipath ceases to exist.

The CBF has significant peaks for multipaths over time [Fig. 12(a)], and the simulated results with the Kraken and the Bellhop match visually well [Figs. 12(b) and 12(d)]. The ray-tracing Bellhop provides the eigen-rays with amplitudes and explains the corresponding multipath.

DOA with dominant strength $\sim -15°$ corresponds to a bottom-reflected (BR) and a surface-bottom-reflected (SBR) path. DOA with a direct path (DP) has the strongest amplitude, but DP and SR 12° are weaker than BR and SBR $-15°$; see time steps 250–350 in Figs. 11(a) and 12(d). As the towed source was near the ocean surface, a path and the path with a surface reflection near the source have a DOA difference of $\sim 1°$ and, thus, are difficult to distinguish. We omit notations for the surface reflected paths, e.g., BR for BR and SBR.

At time steps 220–250, the source is weak at the receiver via DP and SR and BR, and BSR 15° and BSBR $-17°$ dominate. Similarly, multi-SBR paths, e.g., BSBSR and BSBSBR, successively show dominant strength; see time steps 1–200 in Figs. 11(a) and 12(d).

Sparse signal processing, nonsequential, and SVALSE [Figs. 12(c) and 12(e)] result in improved resolution relative to CBF. The CBF [Fig. 12(a)] does not give as high of a resolution as nonsequential VALSE. The SVALSE [Fig. 12(e)] achieves increased accuracy and distinguishes the multipaths; see time steps 1–100.

## IX. CONCLUSION

We introduced an estimation method for sequential direction finding. We target the number of DOAs and their DOAs of time-varying dynamic sources in a statistical model. In the prediction step, sequential processing computes predicted posterior pdfs from the posterior pdfs of the previous time based on a Bernoulli-von Mises state transition model and BP. Variational Bayesian estimation is then used to update predicted posterior pdfs based on a Bernoulli-Gaussian amplitude model that promotes sparsity. Variational Bayesian beamforming in the update step is guaranteed to converge, incorporates prior information, is gridless, and provides the number and posterior pdfs of DOAs. Our method combines BP for sequential processing and MF message passing for variational Bayesian beamforming. We presented sequential variational Bayesian beamforming in a factor graph. The graph visualizes the factorized algorithmic operations and the evolution of DOA information: the propagation of statistical information over time and the update of DOA parameters.

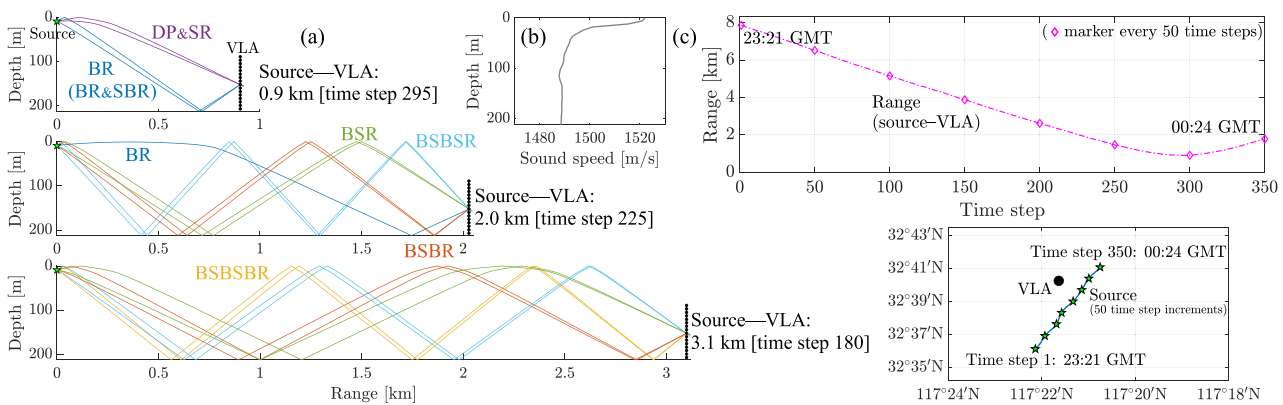

FIG. 11. (Color online) The SWellEx-96 experiment is depicted with (a) multipaths from the VLA and source in range and depth, (b) the water column sound speed from the CTD data, and (c) the range of the VLA source and the track of the source at 23:21–00:24 GMT (time steps 1–350).

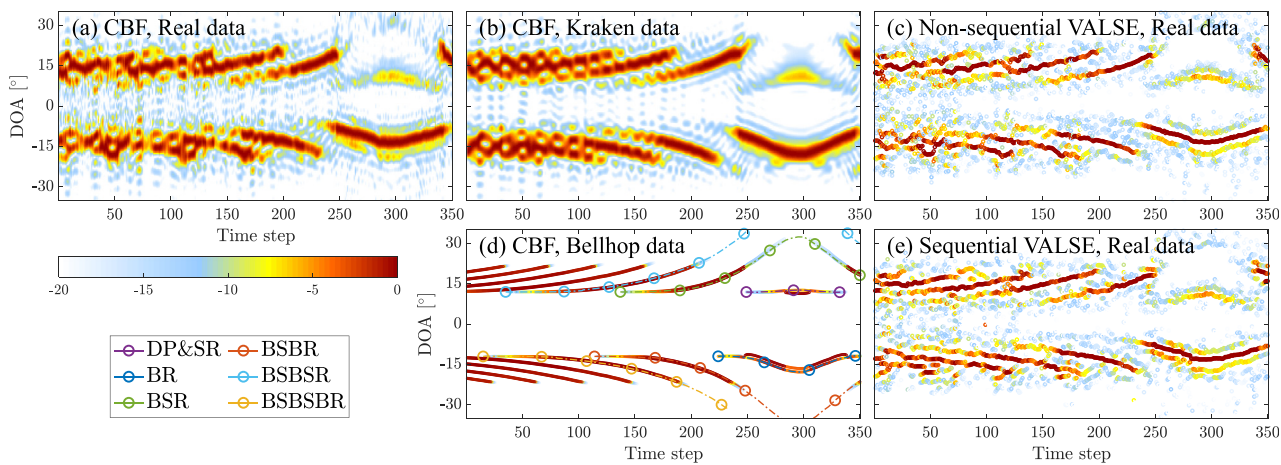

FIG. 12. (Color online) DOA estimation versus time is depicted for (a) CBF, (c) nonsequential VALSE (Refs. 46 and 47), and (e) proposed SVALSE using acoustic data from the SWellEx-96 experiment and CBF using simulated data from (b) normal mode and (d) ray-tracing models.

The simulations indicated that our gridless method estimates the DOAs accurately compared to state-of-the-art techniques. Our method also provides the number of DOAs. We showed the performance with dynamic DOAs that are moving and suddenly (de)activated. The capabilities of our method were demonstrated on experimental ocean acoustics data.

### ACKNOWLEDGMENTS

This research was supported by the Office of Naval Research, Grant No. N00014-21-1-2267. Codes with implementation of the Graph-based sequential beamforming are available at https://github.com/NoiseLabUCSD/SequentialVariationalBayesDOA.

[1]H. L. Van Trees, *Optimum Array Processing* (Wiley, New York, 2002).
[2]H. Krim and M. Viberg, "Two decades of array signal processing research: The parametric approach," IEEE Signal Process. Mag. **13**(4), 67–94 (1996).
[3]D. Malioutov, M. Cetin, and A. S. Willsky, "A sparse signal reconstruction perspective for source localization with sensor arrays," IEEE Trans. Signal Process. **53**(8), 3010–3022 (2005).
[4]D. L. Donoho, "Compressed sensing," IEEE Trans. Inf. Theory **52**(4), 1289–1306 (2006).
[5]E. J. Candès and M. B. Wakin, "An introduction to compressive sampling," IEEE Signal Process. Mag. **25**(2), 21–30 (2008).
[6]P. Gerstoft, C. F. Mecklenbräuker, W. Seong, and M. Bianco, "Introduction to compressive sensing in acoustics," J. Acoust. Soc. Am. **143**(6), 3731–3736 (2018).
[7]A. Xenaki, P. Gerstoft, and K. Mosegaard, "Compressive beamforming," J. Acoust. Soc. Am. **136**(1), 260–271 (2014).
[8]P. Gerstoft, A. Xenaki, and C. F. Mecklenbräuker, "Multiple and single snapshot compressive beamforming," J. Acoust. Soc. Am. **138**(4), 2003–2014 (2015).
[9]D. Shutin and B. H. Fleury, "Sparse variational Bayesian SAGE algorithm with application to the estimation of multipath wireless channels," IEEE Trans. Signal Process. **59**(8), 3609–3623 (2011).
[10]N. Xiang, "Model-based Bayesian analysis in acoustics—A tutorial," J. Acoust. Soc. Am. **148**(2), 1101–1120 (2020).
[11]M. E. Tipping, "Sparse Bayesian learning and the relevance vector machine," J. Mach. Learn. Res. **1**, 211–244 (2001).
[12]D. P. Wipf and B. D. Rao, "Sparse Bayesian learning for basis selection," IEEE Trans. Signal Process. **52**(8), 2153–2164 (2004).
[13]P. Gerstoft, C. F. Mecklenbräuker, A. Xenaki, and S. Nannuru, "Multisnapshot sparse Bayesian learning for DOA," IEEE Signal Process. Lett. **23**(10), 1469–1473 (2016).
[14]K. L. Gemba, S. Nannuru, P. Gerstoft, and W. S. Hodgkiss, "Multi-frequency sparse Bayesian learning for robust matched field processing," J. Acoust. Soc. Am. **141**(5), 3411–3420 (2017).
[15]S. Nannuru, A. Koochakzadeh, K. L. Gemba, P. Pal, and P. Gerstoft, "Sparse Bayesian learning for beamforming using sparse linear arrays," J. Acoust. Soc. Am. **144**(5), 2719–2729 (2018).
[16]G. Ping, E. Fernandez-Grande, P. Gerstoft, and Z. Chu, "Three-dimensional source localization using sparse Bayesian learning on a spherical microphone array," J. Acoust. Soc. Am. **147**(6), 3895–3904 (2020).
[17]H. Niu, P. Gerstoft, E. Ozanich, Z. Li, R. Zhang, Z. Gong, and H. Wang, "Block sparse Bayesian learning for broadband mode extraction in shallow water from a vertical array," J. Acoust. Soc. Am. **147**(6), 3729–3739 (2020).
[18]R. Giri and B. Rao, "Type I and type II Bayesian methods for sparse signal recovery using scale mixtures," IEEE Trans. Signal Process. **64**(13), 3418–3428 (2016).
[19]G. Tang, B. N. Bhaskar, P. Shah, and B. Recht, "Compressed sensing off the grid," IEEE Trans. Inf. Theory **59**(11), 7465–7490 (2013).
[20]E. J. Candès and C. Fernandez-Granda, "Towards a mathematical theory of super-resolution," Commun. Pure Appl. Math. **67**(6), 906–956 (2014).
[21]P. Pal and P. P. Vaidyanathan, "A grid-less approach to underdetermined direction of arrival estimation via low rank matrix denoising," IEEE Signal Process. Lett. **21**(6), 737–741 (2014).
[22]Y. Chi and M. Ferreira Da Costa, "Harnessing sparsity over the continuum: Atomic norm minimization for superresolution," IEEE Signal Process. Mag. **37**(2), 39–57 (2020).
[23]C. F. Mecklenbräuker, P. Gerstoft, A. Panahi, and M. Viberg, "Sequential Bayesian sparse signal reconstruction using array data," IEEE Trans. Signal Process. **61**(24), 6344–6354 (2013).
[24]F. Meyer, Y. Park, and P. Gerstoft, "Variational Bayesian estimation of time-varying DOAs," in *Proceedings of IEEE FUSION* (2020), pp. 1–6.
[25]Y. Park, F. Meyer, and P. Gerstoft, "Sequential sparse Bayesian learning for time-varying direction of arrival," J. Acoust. Soc. Am. **149**(3), 2089–2099 (2021).
[26]Y. Park and P. Gerstoft, "Gridless sparse covariance-based beamforming via alternating projections including co-prime arrays," J. Acoust. Soc. Am. **151**(6), 3828–3837 (2022).
[27]Y. Park, P. Gerstoft, and J. H. Lee, "Difference-frequency MUSIC for DOAs," IEEE Signal Process. Lett. **29**, 2612–2616 (2022).
[28]B. Ristic, M. S. Arulampalam, and N. Gordon, *Beyond the Kalman Filter: Particle Filters for Tracking Applications* (Artech House, Norwood, MA, 2004).
[29]Y. Bar-Shalom, P. K. Willett, and X. Tian, *Tracking and Data Fusion: A Handbook of Algorithms* (Bar-Shalom, Storrs, CT, 2011).

[30]C. Yardim, Z. H. Michalopoulou, and P. Gerstoft, "An overview of sequential Bayesian filtering in ocean acoustics," IEEE J. Ocean. Eng. **36**(1), 71–89 (2011).

[31]J. Li and H. Zhou, "Tracking of time-evolving sound speed profiles in shallow water using an ensemble Kalman-particle filter," J. Acoust. Soc. Am. **133**(3), 1377–1386 (2013).

[32]J. V. Candy, "Environmentally adaptive processing for shallow ocean applications: A sequential Bayesian approach," J. Acoust. Soc. Am. **138**(3), 1268–1281 (2015).

[33]A. Simonetto, E. Dall'Anese, S. Paternain, G. Leus, and G. B. Giannakis, "Time-varying convex optimization: Time-structured algorithms and applications," Proc. IEEE **108**(11), 2032–2048 (2020).

[34]X. Li, E. Leitinger, and F. Tufvesson, "Detection and tracking of multipath channel parameters using belief propagation," in *Proceedings of the Asilomar Conference on Signals, Systems & Computers (ACSSC)* (2020), pp. 1083–1089.

[35]X. Li, E. Leitinger, A. Venus, and F. Tufvesson, "Sequential detection and estimation of multipath channel parameters using belief propagation," IEEE Trans. Wireless Commun. **21**(10), 8385–8402 (2022).

[36]J. Ziniel and P. Schniter, "Dynamic compressive sensing of time-varying signals via approximate message passing," IEEE Trans. Signal Process. **61**(21), 5270–5284 (2013).

[37]R. Prasad, C. R. Murthy, and B. D. Rao, "Joint approximately sparse channel estimation and data detection in OFDM systems using sparse Bayesian learning," IEEE Trans. Signal Process. **62**(14), 3591–3603 (2014).

[38]M. S. Asif and J. Romberg, "Sparse recovery of streaming signals using $\mathcal{L}_1$-homotopy," IEEE Trans. Signal Process. **62**(16), 4209–4223 (2014).

[39]A. S. Charles, A. Balavoine, and C. J. Rozell, "Dynamic filtering of time-varying sparse signals via $\mathcal{L}_1$ minimization," IEEE Trans. Signal Process. **64**(21), 5644–5656 (2016).

[40]N. Vaswani and J. Zhan, "Recursive recovery of sparse signal sequences from compressive measurements: A review," IEEE Trans. Signal Process. **64**(13), 3523–3549 (2016).

[41]M. R. O'Shaughnessy, M. A. Davenport, and C. J. Rozell, "Sparse Bayesian learning with dynamic filtering for inference of time-varying sparse signals," IEEE Trans. Signal Process. **68**, 388–403 (2020).

[42]Y. Yang, Z. Chu, Z. Xu, and G. Ping, "Two-dimensional grid-free compressive beamforming," J. Acoust. Soc. Am. **142**(2), 618–629 (2017).

[43]T. L. Hansen, B. H. Fleury, and B. D. Rao, "Superfast line spectral estimation," IEEE Trans. Signal Process. **66**(10), 2511–2526 (2018).

[44]Y. Park, Y. Choo, and W. Seong, "Multiple snapshot grid free compressive beamforming," J. Acoust. Soc. Am. **143**(6), 3849–3859 (2018).

[45]G. Chardon and U. Boureau, "Gridless three-dimensional compressive beamforming with the Sliding Frank-Wolfe algorithm," J. Acoust. Soc. Am. **150**(4), 3139–3148 (2021).

[46]M.-A. Badiu, T. L. Hansen, and B. H. Fleury, "Variational Bayesian inference of line spectra," IEEE Trans. Signal Process. **65**(9), 2247–2261 (2017).

[47]J. Zhu, Q. Zhang, P. Gerstoft, M.-A. Badiu, and Z. Xu, "Grid-less variational Bayesian line spectral estimation with multiple measurement vectors," Signal Process. **161**, 155–164 (2019).

[48]J. Zhu, C. K. Wen, J. Tong, C. Xu, and S. Jin, "Grid-less variational Bayesian channel estimation for antenna array systems with low resolution ADCs," IEEE Trans. Wireless Commun. **19**(3), 1549–1562 (2020).

[49]Q. Zhang, J. Zhu, N. Zhang, and Z. Xu, "Multidimensional variational line spectra estimation," IEEE Signal Process. Lett. **27**, 945–949 (2020).

[50]C. M. Bishop, *Pattern Recognition and Machine Learning* (Springer, Berlin, 2006).

[51]K. P. Murphy, *Machine Learning: A Probabilistic Perspective* (MIT Press, Cambridge, MA, 2012).

[52]K. P. Murphy, *Probabilistic Machine Learning: Advanced Topics* (MIT Press, Cambridge, MA, 2023).

[53]M. J. Wainwright and M. I. Jordan, "Graphical models, exponential families, and variational inference," TR-649, University of California, Berkeley (2003).

[54]T. Minka, "Divergence measures and message passing," MSR-TR-2005-173, Microsoft Research (2005).

[55]B. Cakmak, D. N. Urup, F. Meyer, T. Pedersen, B. H. Fleury, and F. Hlawatsch, "Cooperative localization for mobile networks: A distributed belief propagation-mean field message passing algorithm," IEEE Signal Process. Lett. **23**, 828–832 (2016).

[56]F. R. Kschischang, B. J. Frey, and H.-A. Loeliger, "Factor graphs and the sum-product algorithm," IEEE Trans. Inf. Theory **47**(2), 498–519 (2001).

[57]A. T. Ihler, J. W. Fisher III, R. L. Moses, and A. S. Willsky, "Nonparametric belief propagation for self-localization of sensor networks," IEEE J. Sel. Areas Commun **23**(4), 809–819 (2005).

[58]F. Meyer, O. Hlinka, H. Wymeersch, E. Riegler, and F. Hlawatsch, "Distributed localization and tracking of mobile networks including non-cooperative objects," IEEE Trans. Signal Inf. Process. Netw. **2**(1), 57–71 (2016).

[59]F. Meyer, T. Kropfreiter, J. L. Williams, R. A. Lau, F. Hlawatsch, P. Braca, and M. Z. Win, "Message passing algorithms for scalable multitarget tracking," Proc. IEEE **106**(2), 221–259 (2018).

[60]H. Naseri and V. Koivunen, "A Bayesian algorithm for distributed network localization using distance and direction data," IEEE Trans. Signal Inf. Process. Netw. **5**(2), 290–304 (2019).

[61]F. Meyer and J. L. Williams, "Scalable detection and tracking of geometric extended objects," IEEE Trans. Signal Process. **69**, 6283–6298 (2021).

[62]J. Jang, F. Meyer, E. R. Snyder, S. M. Wiggins, S. Baumann-Pickering, and J. A. Hildebrand, "Bayesian detection and tracking of odontocetes in 3-D from their echolocation clicks," available at https://arxiv.org/abs/2210.12318 (Last viewed January 22, 2023).

[63]E. Riegler, G. E. Kirkelund, C. N. Manchon, M.-A. Badiu, and B. H. Fleury, "Merging belief propagation and the mean field approximation: A free energy approach," IEEE Trans. Inf. Theory **59**(1), 588–602 (2013).

[64]K. V. Mardia and P. E. Jupp, *Directional Statistics* (Wiley, Chichester, UK, 2000).

[65]J. Dauwels, "On variational message passing on factor graphs," in *Proceedings of the IEEE International Symposium on Information Theory (ISIT)* (2007), pp. 2546–2550.

[66]H. V. Poor, *An Introduction to Signal Detection and Estimation*, 2nd ed. (Springer, New York, 1994).

[67]K. P. Murphy, *Probabilistic Machine Learning: An Introduction* (MIT Press, Cambridge, MA, 2022).

[68]D. M. Blei, A. Kucukelbir, and J. D. McAuliffe, "Variational inference: A review for statisticians," J. Am. Stat. Assoc. **112**(518), 859–877 (2017).

[69]D. P. Kingma and M. Welling, "Auto-encoding variational Bayes," in *Proceedings of the International Conference on Learning Representations* (2014), pp. 1–14.

[70]A. S. Rahmathullah, Á. F. García-Fernández, and L. Svensson, "Generalized optimal sub-pattern assignment metric," in *Proceedings of the IEEE International Conference on Information Fusion* (2017), pp. 1–8.

[71]D. Schuhmacher, B.-T. Vo, and B.-N. Vo, "A consistent metric for performance evaluation of multi-object filters," IEEE Trans. Signal Process. **56**(8), 3447–3457 (2008).

[72]K. L. Gemba, S. Nannuru, and P. Gerstoft, "Robust ocean acoustic localization with sparse Bayesian learning," IEEE J. Sel. Top. Signal Process. **13**(1), 49–60 (2019).

[73]F. Meyer and K. L. Gemba, "Probabilistic focalization for shallow water localization," J. Acoust. Soc. Am. **150**(2), 1057–1066 (2021).

[74]M. B. Porter, "Acoustic toolbox," available at http://oalib.hlsresearch.com/AcousticsToolbox/ (Last viewed January 22, 2023).

[75]G. L. D'Spain, J. J. Murray, W. S. Hodgkiss, N. O. Booth, and P. W. Schey, "Mirages in shallow water matched field processing," J. Acoust. Soc. Am. **105**(6), 3245–3265 (1999).